\documentclass[twocolumn]{emulateapj}
\usepackage{xcolor}
\usepackage{ulem}
\usepackage{lineno}

\usepackage{multirow}
\usepackage{tabularx}

\slugcomment{Not to appear in Nonlearned J., 45.}

\shorttitle{Weak temporal variations in the solar emission}
\shortauthors{Sharma et al.}

\begin{document}


\title{Detection of weak ubiquitous impulsive nonthermal emissions from the solar corona}


\author{Rohit Sharma\altaffilmark{1}, Divya Oberoi\altaffilmark{2}, Marina Battaglia\altaffilmark{1}, and S{\"a}m Krucker\altaffilmark{1}}
\affil{$^1$ Fachhochschule Nordwestschweiz,
Bahnhofstrasse 6, 5210 Windisch, Switzerland}
\affil{$^2$National Centre for Radio Astrophysics, Tata Institute of Fundamental Reseach, Pune 411020, India}
\email{rohit.sharma@fhnw.ch}



\begin{abstract}
A ubiquitous presence of weak energy releases is one of the most promising hypotheses to explain coronal heating, referred to as the nanoflare hypothesis.
The accelerated electrons associated with such weak heating events are also expected to give rise to coherent impulsive emission via plasma instabilities in the  meterwave radio band, making this a promising spectral window to look for their presence.
Recently \citet{Mondal2020b} reported the presence of weak impulsive emissions from quiet Sun regions which seem to meet the requirements of being radio counterparts of the hypothesized nanoflares.
Detection of such low-contrast weak emission from the quiet Sun is challenging and, given their implications, it is important to confirm their presence. 
In this work, using data from the Murchison Widefield Array, we explore the use of an independent robust approach for their detection by separating the dominant slowly varying component of emission from the weak impulsive one in the visibility domain.
We detect milli-SFU level bursts taking place all over the Sun and characterize their brightness temperatures, distributions, morphologies, durations and association with features seen in EUV images.
We also attempt to constraint the energies of the nonthermal particles using inputs from the FORWARD coronal model along with some reasonable assumptions and find them to lie in the sub-pico flare ($\sim 10^{19}-10^{21}$ ergs) range. 
In the process, we also discover perhaps the weakest type III radio burst and another one that shows clear signatures of weakest quasi-periodic pulsations.
\end{abstract}

\section{Introduction}

Solar radio bursts are valuable probes of the nonthermal particle populations produced during energy releases in the solar corona. The accelerated particles interact with the ambient plasma and, via plasma instabilities, produce a variety of electromagnetic emissions observed at radio wavelengths. 
The brightest radio emissions, the type-III bursts, are impulsive and are produced by supra-thermal electron beams streaming across the solar corona, instigating a bump-on-the-tail instability \citep[e.g.][]{Reid2014, Sharma2017, Reid2020}. The most common radio emission, type-I bursts, are relatively faint, can last for hours to days, and have been linked to emerging magnetic flux via loss-cone instability \citep{Spicer1982}. 
The amount of energy released depends on the nature of the instability and ambient local coronal parameters like magnetic field strengths, electron temperature, and densities. A wide variety of active solar emissions differing vastly in their spectral and temporal characteristics have been known since the 1950s and have been characterized in great detail \citep[e.g.][]{McLean1985}. 

With the availability of increasingly sensitive and higher temporal and spectral resolution radio measurements from the new generation of radio interferometers and the much improved imaging characteristics they provide, solar emissions are being explored in ever greater detail. 
Recent examples include -- finding previously unknown finer emission features in well known types of burst emissions e.g. \cite{Jasmina2020} presents fine structures in a Type II and \cite{Reid2021} in a type III; discovery of quasi-periodic pulsations in a large variety of solar active emissions \citep{Mohan2019}; and observations of faint gyrosychrotron emission becoming more common (\cite{Mondal2019}).
In particular, the solar observations from the Murchison Widefield Array \citep[MWA;][]{Lonsdale2009,Tingay2013} have lead to an increasing awareness of progressively weaker nonthermal radio bursts present even during relatively quiet times.
The earlier such studies relied on non-imaging techniques \citep[][]{Suresh2017, Sharma2018}.
Once a robust analysis pipeline for high fidelity and dynamic range snapshot spectroscopic imaging capability became available \citep{Mondal2019}, more recent studies are based on radio images from this pipeline. 
The most interesting work, in the present context, is by \citet{Mondal2020b}, which presents convincing evidence for weak impulsive narrowband emissions at milli-SFU (mSFU; 1 SFU = 10$^4$ Jy) level from the quiet Sun regions.
Estimates for the energy deposited in the corona by groups of such emissions lies close to the nanoflare regime \citep{Mondal2021SoPh}.
There is, hence, a promising possibility that these could be the radio counterparts of the so called ``nanoflares'', the small energy releases  originally hypothesises by \citet{Parker1988} as a potential resolution to the coronal heating puzzle.
As nonthermal meterwave solar emissions originate via coherent emission, they are ideal probes for energetically weaker events than can be studied at EUV and X-ray bands.

\citet{Mondal2020b} identified the weak impulsive bursts by looking for outliers in the radio light curves from many small regions distributed all over the Sun, each having an angular size of the instrumental resolution.
One of the key next steps in understanding these bursts better is to image them.
It is, however, challenging to locate and image these weak short-lived low-contrast bursts as their strengths are of order 1\% of the thermal bremsstrahlung dominated background emission.
Passage through coronal inhomogenities also leads to significant scatter broadening which further reduces the contrast, an especially significant  effect for emission at the fundamental or even the harmonic of the local plasma frequency \citep[e.g.][]{Kontar2019,Chrysaphi2018, Sharma2020, Mohan2021b}. 

In this work we demonstrate the efficacy of a technique for removal of slowly varying thermal background emission in the localization and imaging of these weak transient emissions.
The background emission is estimated using the median of the data over a short running window and removed by vector subtraction in the visibility domain, individually for each baseline.
By avoiding deconvolution errors from bright static background incurred during the imaging process, this approach offers a significant advantage over standard imaging.
Similar techniques based on visibility subtraction have also been used earlier to study transient burst emissions \citep[e.g.][]{Marsh1980,Sharma2020M,Luo2021,Ryan2021}.
We provide a detailed characterization of the nature and distribution of these burst emissions, estimates of their energies and discuss their implications for coronal heating.

The paper begins by providing a description of the MWA solar data and imaging analysis in Section \ref{sec:observation}. In Section \ref{sec:radio-analysis} we describe the visibility subtraction technique, while Section \ref{sec:burst-characterisation} presents a characterization of various aspects of these weak bursts.
The aspects related to energetics of these bursts are described in Section \ref{sec:energetics}  
followed by the discussion and conclusion in Sections \ref{sec:discussion} and \ref{sec:conclusion} respectively.

\section{Observations} \label{sec:observation}
The Murchison Widefield Array (MWA) is a precursor of the Square Kilometer Array (SKA), located in Western Australia. It is a new-generation ``large-N" radio interferometer operating in the 80-300 MHz frequency range \citep{Lonsdale2009,Tingay2013}. We use solar MWA data from the phase-I configuration which has 128 elements spread over a region of $\sim$3 km in diameter in a centrally condensed configuration. 
The dense UV coverage due to the large number of short baselines provided by the MWA yield high fidelity and sensitive snapshot spectroscopic radio imaging of the quiet Sun of unprecedented quality.

We use the MWA solar data from 03 December 2015, acquired under the project code G0002. According to {\tt www.solarmoniter.org}, the level of solar activity was low on this day. 
Assuming the Newkirk density model \citep{Newkirk1961}, the observed MWA frequencies sample the coronal height range from 15 to 150 Mm above the photosphere. 
The Extreme-Ultra Violet (EUV) maps from AIA onboard SDO \citep{Pesnell2012} show presence of a large coronal hole on the solar disk (Fig. \ref{Fig:intro1} (A)). Coronal Hole Identification via multi-thermal Emission Recognition Algorithm \citep[CHIMERA;][]{Garton2018} quantified the coronal hole area as 23\% of the solar disk. 
The coronal hole largely lies in the northern hemisphere, but does cross the equator and enters into the southern hemisphere. A smaller coronal hole is also present at the south pole.
Figure \ref{Fig:intro1} (A) also shows the composite AIA map at 171, 193 and 211 \AA\ wavelengths. 
Most of the EUV bright regions are located on the limb. 
 Figure \ref{Fig:intro1}(B) shows the HMI magnetogram.
All of the NOAA classified active regions are marked and are seen to lie close to the western limb. 
The HMI magnetogram also shows emerging polarities on the western, eastern and central sections, i.e. at about (600", 100"), (-700", 70") and (-100",350") respectively. 
 All of them have EUV counter-parts associated with them.
GOES X-ray light curves are essentially featureless during the time of radio observations marked by shaded blue region in the Fig. \ref{Fig:intro2}(A). 
However after the radio observations analyzed here, two C-class flares occurred. The C2.2 and C1.2 flares happened in the AR2458 on the western limb \ref{Fig:intro1}(B) at 04:46:00 UT and 06:01:00 UT respectively. A corresponding X-ray flux increase can be seen at 04:50 UT for the first flare, but only second flare is listed in Space Weather Prediction Center (SWPC\footnote{https://www.swpc.noaa.gov}). 



\section{Radio Data Analysis} \label{sec:radio-analysis}
We analyse a total of $\sim$30 minutes of MWA solar data obtained on 03 December 2015 from 03:24 UT to 03:40 UT and from 03:44 UT to 03:59 UT. These observations were taken in the so called ``picket-fence" mode often used for the Sun, where the total bandwidth of 30.72 MHz was divided into twelve 2.56 MHz wide coarse bands distributed in a roughly log-spaced manner from 80 to 240 MHz. 
We present the analysis of 8 of these frequency bands, the ones centered roughly at 108, 120, 133, 145, 179, 197, 217 and 240 MHz.

\subsection{Estimating the total Solar Flux Density}
\begin{figure*}
    \centering
\resizebox{170mm}{!}{
\includegraphics[trim={0.0cm 0cm 0.0cm 0.0cm},clip,scale=0.3]{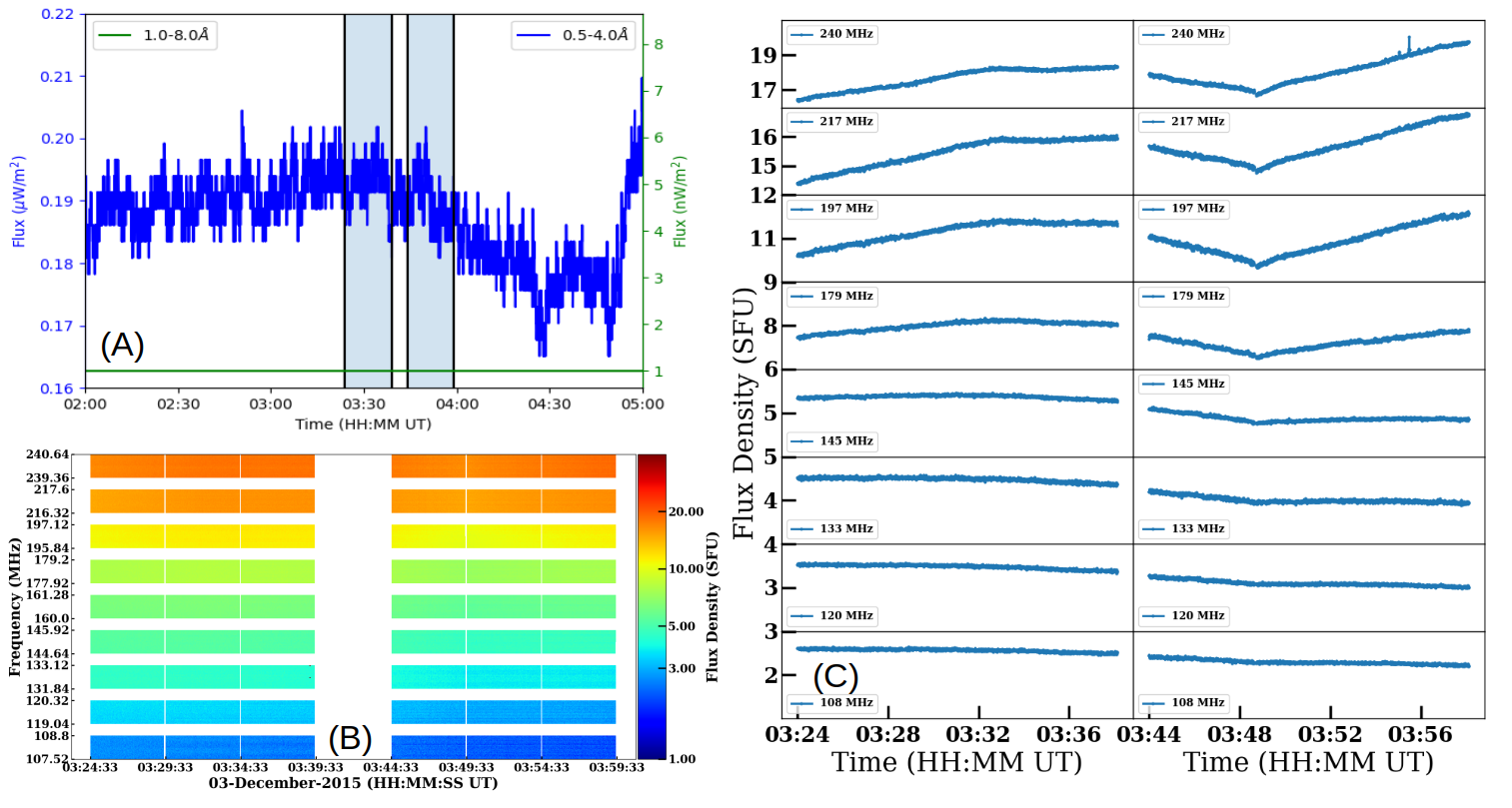}}
    \caption{ Panel (A): GOES X-ray lightcurves. The shaded region corresponds to the times overlapped with the radio observations. Panel (B): Dynamic Spectrum of the solar-disk integrated flux densities for the two 15 min datasets. Due to limitations in data volumes, we omit the middle 5 min of data in the analysis. Panel (C): Frequency-averaged timeseries of the flux densities for the eight frequency bands. Note that the time intervals between left and right columns are different.}
    \label{Fig:intro2}
\end{figure*}

The solar flux density at meterwaves is known to vary over timescales of minutes. This has historically been referred to as the slowly varying component of solar flux density \citep[e.g.][]{McLean1985,Sharma2018}. 
We compute the disk-integrated solar flux at desired eight frequency bands using the prescription given by \cite{Oberoi2017}. 
Figure \ref{Fig:intro2} (B) shows the flux calibrated Dynamic Spectrum (DS) and Fig. \ref{Fig:intro2} (C), the spectrally averaged light curve for each of the eight frequency bands. 
The time series show a low level smoothly and slowly varying component of $\approx$ 10\% for all frequencies. 
The variation is a little larger at higher frequencies, as compared to low frequencies. 
Interestingly, the 240 MHz frequency band shows a few faint impulsive bursts around 03:55:30 UT, the brightest of which is $\sim$0.5 SFU.

\begin{figure*}
    \centering
    \begin{tabular}{cc}
    \resizebox{70mm}{!}{
\includegraphics[trim={0.0cm 0cm 0.0cm 0.0cm},clip,scale=0.3]{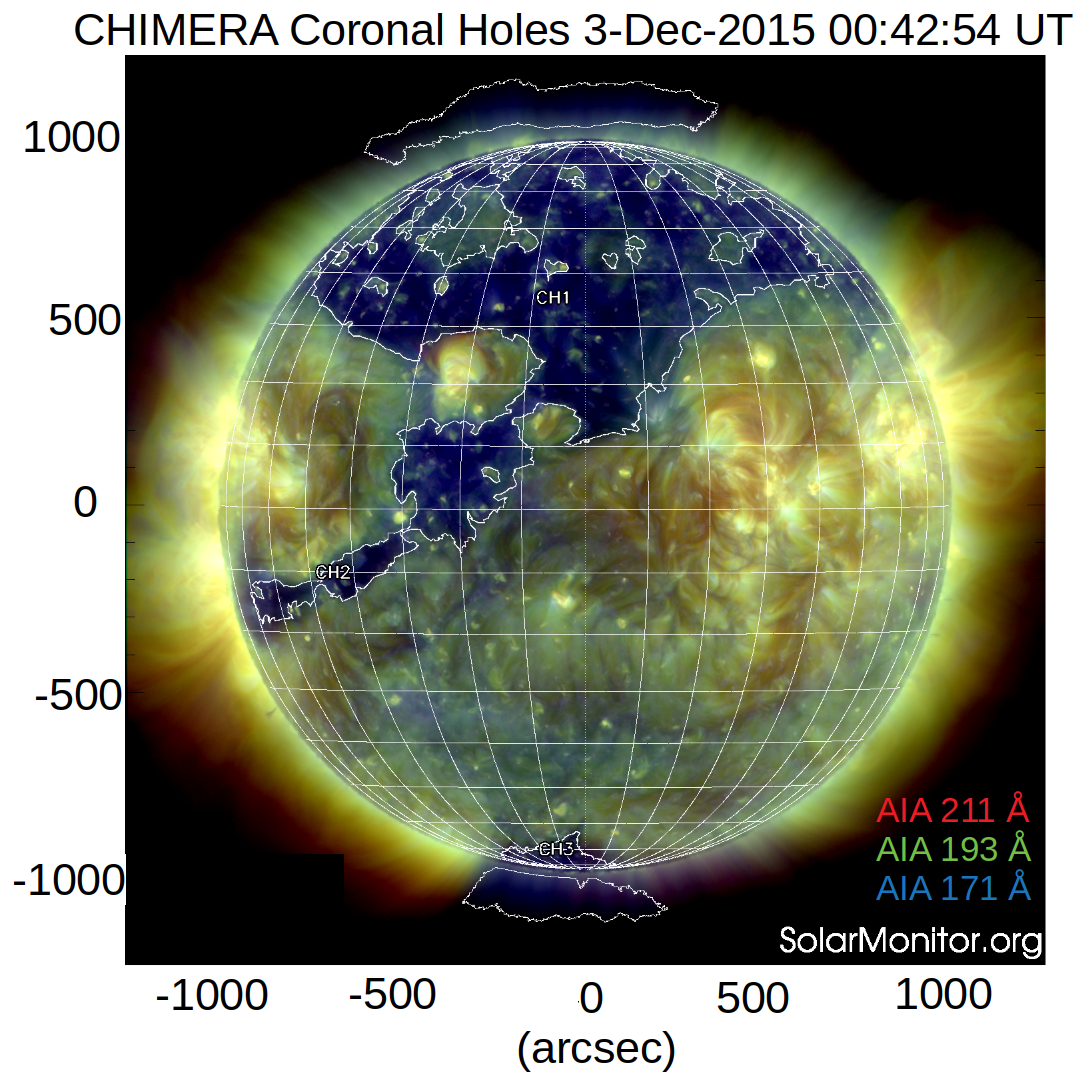}}
         &    \resizebox{70mm}{!}{
\includegraphics[trim={0.0cm 0cm 0.0cm 0.0cm},clip,scale=0.3]{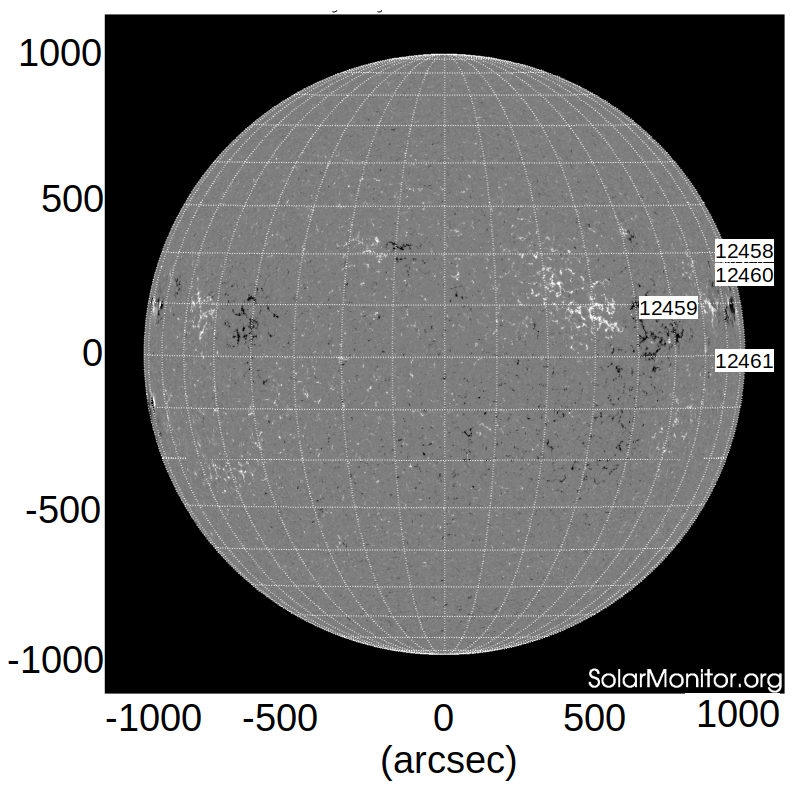}}  \\
       (A) AIA 193 \AA\  & (B) HMI magnetogram
    \end{tabular}
    \caption{Panel (A): Composite EUV images for 171 \AA\ (blue), 193 \AA\ (green), and 211 \AA\ (red). CHIMERA coronal hole boundary is shown as white contours. Panel (B): HMI magnetogram at 18:46:09 UT}
    \label{Fig:intro1}
\end{figure*}

\begin{table}
    \centering
    
    \begin{tabular}{|c|c|c|}
    \hline
      Frequency (MHz)   & B$_{maj}$ & B$_{min}$ \\
      \hline
        108 & 3.71" & 3.51" \\
        120 & 3.50" & 3.33"\\
        133 & 3.39" & 3.23" \\
        145 & 3.35" & 3.18"\\
        179 & 3.21" & 3.07" \\
        197 & 3.17" & 3.09"\\
        217 & 3.13" & 3.10"\\
        240 & 3.15" & 3.08"\\
            \hline
    \end{tabular}
    \caption{The major and minor axis of the MWA PSF for various frequencies.}
    \label{Tab:psf}
\end{table}

\subsection{Radio Imaging of the Sun}

We perform radio imaging of the solar disk for the eight frequency bands (108 to 240 MHz) using CASA \citep{McMullin2007}. The images are produced at 0.5 s time-resolution and averaged over the $\sim$2 MHz of usable bandwidth for each of the frequency bands.
Typical dynamic range of the images lies between 100 to 200.
The solar disk boundary was chosen to lie at 5$\sigma_{map}$, where $\sigma_{map}$ is the RMS computed over a large region in the image away from the Sun. 
These images were converted into brightness temperature ($T_B$) maps following the prescription given by \citet{Mohan2017}.

\begin{figure*}
\begin{center}
\begin{tabular}{c}
\resizebox{150mm}{!}{
\includegraphics[trim={0.0cm 0cm 0.0cm 0.0cm},clip,scale=0.3]{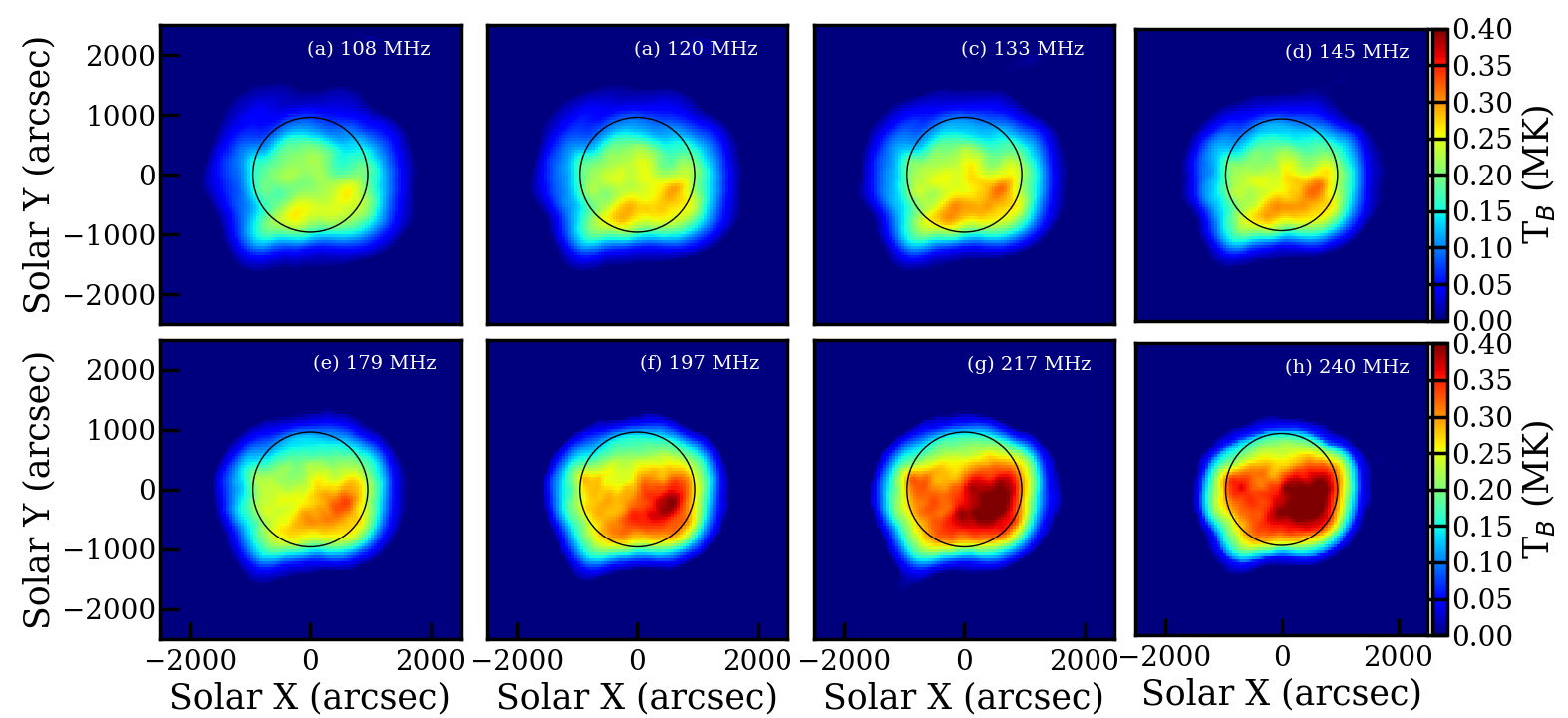}}
\end{tabular}
\end{center}
\caption{ Radio solar maps for the eight frequency bands. 
The colorbar is saturated at 0.4 MK. The peak brightness temperature from 108 MHz to 240 MHz are 0.26, 0.30, 0.32, 0.34, 0.37, 0.40, 0.44 and 0.47 MK respectively.
The black circle in each panel marks the size of the visible Sun, i.e. 16' radius. 
The maps were time-averaged over the entire 30 min ranging from 03:24 to 03:39 UT and 03:44 to 03:59 UT.
\label{Fig:maps}}
\end{figure*}

Figure \ref{Fig:maps} shows the time-averaged $T_B$ solar maps. The peak $T_B$ drops roughly by half from $\sim$0.47 MK to $\sim$0.27MK, and a clear increase in disk size is observed as a function of frequency from 240 MHz to 108 MHz. 
During quiet times, solar radio emission at metrewaves is produced  pre-dominantly via thermal bremsstrahlung in the optically thin corona. In addition to the intrinsic variations in the coronal medium, propagation effects through the turbulent and inhomogeneous corona, like refraction and scattering, also contribute to the observed differences in solar radio maps across frequencies \citep{Sharma2020}.
Overall, the good correspondence between the on-disk EUV bright features and radio bursts is clearly seen at higher radio frequencies (Figs. \ref{Fig:intro1}(A) and \ref{Fig:maps}).
The coronal hole region transitions from being a radio dark region (i.e. a region of lower $T_B$ as compared to neighboring regions) at higher radio frequencies, to being a radio bright region at lower radio frequencies (e.g. 120 MHz). 
Such transitions of $T_B$ of coronal holes have been observed earlier as well \citep{Rahman2019}.

\section{Characterizing weak bursts}
\label{sec:burst-characterisation}

The observed solar emission can be modeled as the superposition of a temporally slowly varying component and impulsive bursts \citep{Sharma2018}.
In this work, we refer to all impulsive radio emissions as bursts, independent of their strengths.
This section describes our approach to isolating these weak bursts from the total solar radio emission, the image noise characteristics, and the duration of the bursts.

\subsection{Modeling observed visibilities} \label{sec:subvs}

Let $\vec{V}(u,v,w,\nu,t)$ be the calibrated cross-correlation or visibility for the baseline $(u,v,w)$ at frequency $\nu$ and time $t$. 
Though the intrinsic solar emission varies across each of these parameters, in the following treatment we deal with visibility light curves only one baseline and frequency at a time.
So we simplify our notation to only emphasise the temporal variability aspect i.e. $\vec{V}(u,v,w,\nu,t) \equiv \vec{V}(t)$.
The thermal emission is expected to be spectrally smooth and, as it arises from large scale coronal electron populations, its temporal variations are also expected to be slow, consistent with large-scale variations in the coronal plasma distribution \citep[e.g.][]{Sharma2018}.

Hence, we model the observed visibilities as the sum of two parts, a slowly varying part, $\vec{V}_{s}(t)$, corresponding primarily to the thermal emission component, and an impulsive part, $\vec{V}_{i}(t)$, which arises primarily from the nonthermal active emission processes.

\begin{equation}
\vec{V}(t) = \vec{V}_{s}(t) + \vec{V}_{i}(t)
\end{equation}

\begin{figure*}
\begin{center}
\begin{tabular}{ccc}
\resizebox{54mm}{!}{
\includegraphics[trim={0.0cm 0cm 0.0cm 0.0cm},clip,scale=0.3]{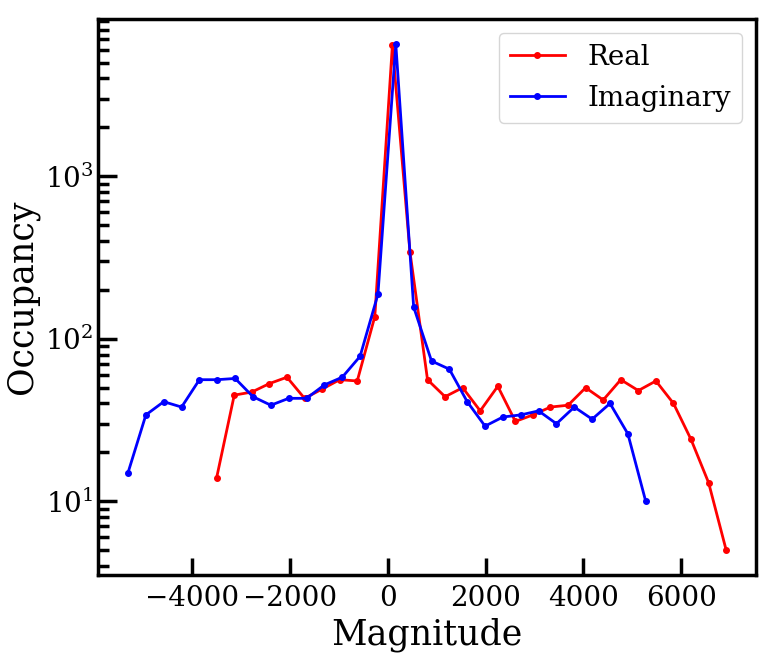}} &
\resizebox{54mm}{!}{
\includegraphics[trim={0.0cm 0cm 0.0cm 0.0cm},clip,scale=0.3]{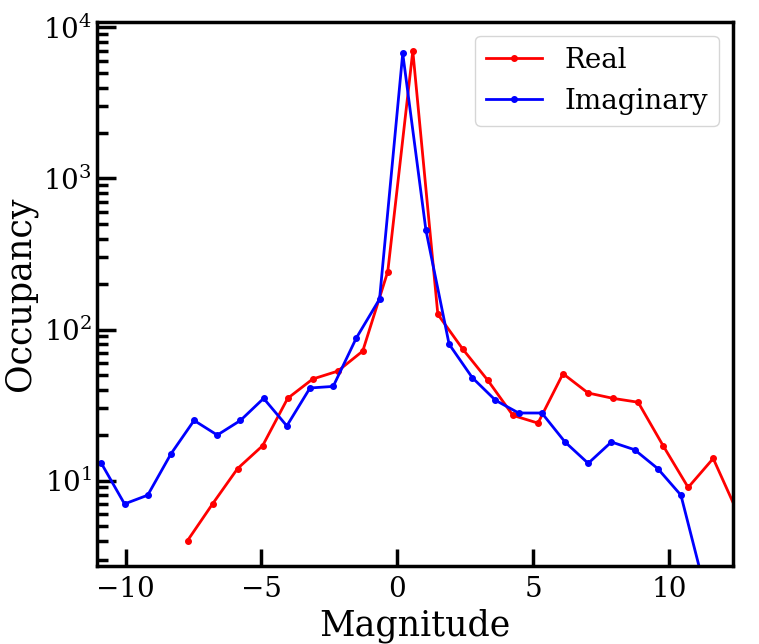}} &
\resizebox{60mm}{!}{
\includegraphics[trim={0.0cm 0cm 0.0cm 0.0cm},clip,scale=0.3]{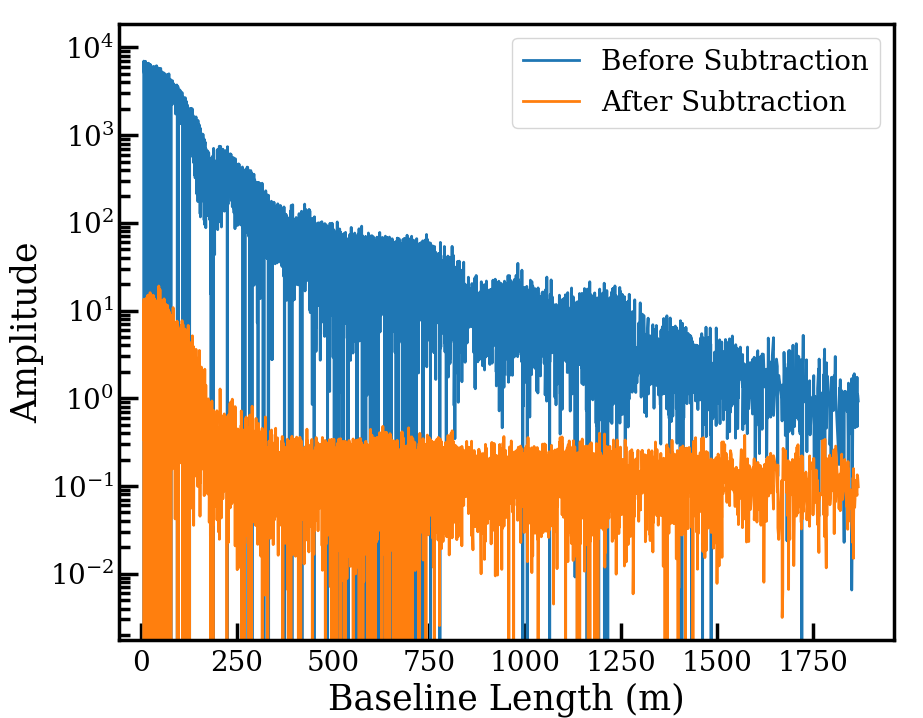}}\\

(A) Before subtraction (1$\sigma$ width = 3200) & (B) After subtraction (1$\sigma$ width = 8) & (C) Visibility amplitudes before and after\\
 & & subtraction

\end{tabular}
\end{center}
\caption{ Panel A \ and  B shows the histogram of the real and imaginary parts of the complex visibilities before and after continuum subtraction at 179 MHz. Panel C shows the visibility amplitude before and after the subtraction versus baseline length for 179 MHz.
\label{Fig:vis_hist}}
\end{figure*}

As these data come from a rather quiet time, $\vec{V}_{i}(t)$ is expected to be much smaller than $\vec{V}_{s}(t)$.
To separate out the low level $\vec{V}_{i}(t)$ from the observed $\vec{V}(t)$, at any given frequency, we work with light curves from one baseline at a time and independently subtract a running median from the light curves of real and imaginary parts of $\vec{V}(t)$, respectively. 
The running median was computed over a 15 s window, and outliers beyond 5$\sigma_{map}$ were removed from the light curve prior to median computation.
The histograms of real and imaginary parts of $\vec{V}(t)$ and $\vec{V}_i(t)$ are shown in the Fig. \ref{Fig:vis_hist} (A \& B).
Noting the spans on the x-axes, it is evident that the running median subtraction has reduced these numbers by $\sim$99.75\%.
The Fig. \ref{Fig:vis_hist} (C) shows the visibility amplitude as a function of baseline length before and after subtraction.
It is evident that the contributions to shorter baselines is reduce by about three orders of magnitude, while those to the longest baselines are reduced by about an order of magnitude.
This is consistent with the expectation of the thermal coronal emission dominating the emission at large angular scales (short baselines) and the contribution of weak nonthermal emissions, believed to arise primarily from compact sources, contributing a larger fraction of the observed visibilities at longer baselines.

\subsection{Imaging using residual visibilities}

A series of maps with a time resolution of 0.5 s were made for each of the same eight frequency bands, this time using the impulsive component of visibilities, $V_i(t)$. 
We refer to these images as {\it residual images}.
The noise levels in a part of the map away from the Sun were examined to look for outliers. 
At any given frequency, the noise levels were stable across time for most of the images, though a few time slices ($\sim 8\%$) showed abrupt jumps and were excluded from further analysis.
For each frequency, the $T_B$ residual maps were computed using the method described by \citet{Mohan2017}.

\begin{figure}
    \centering
    \resizebox{80mm}{!}{
\includegraphics[trim={0.0cm 0cm 0.0cm 0.0cm},clip,scale=0.3]{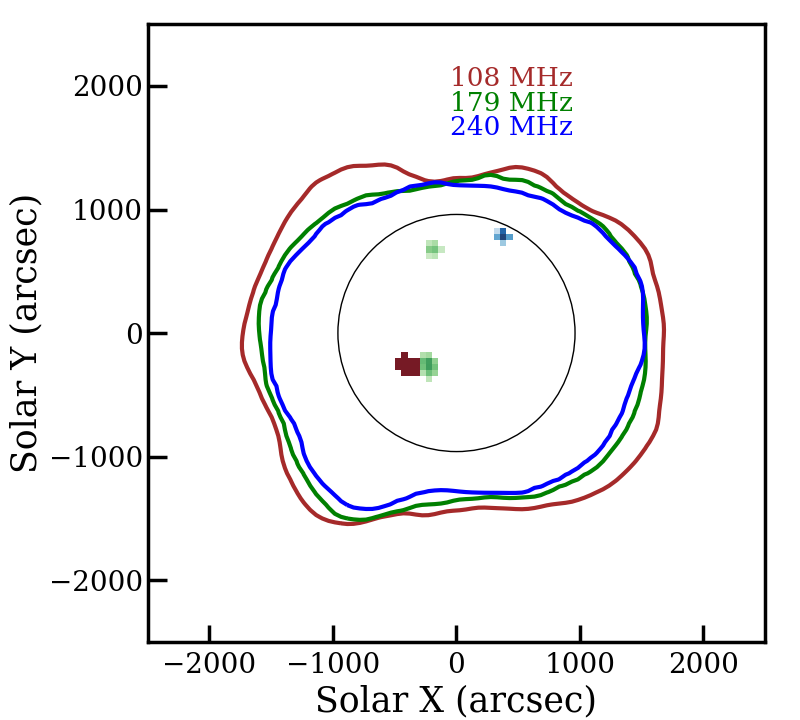}}
    \caption{Examples of transient events seen in the residual solar maps. The black disk marks the size of the optical Sun. The contours indicate the boundary of the radio Sun by marking a level of 2$\times 10^4$ K. These maps come from the time range 03:24:36.5 to 03:24:37.0 UT.
    }
    \label{Fig:example}
\end{figure}

The most prominent features in the residual maps are compact sources which appear at varying locations on the solar disc.
Figure \ref{Fig:example} shows radio images at three different frequencies for a randomly chosen 0.5 s time slice, as an example of the detected compact burst.
One or more compact bursts are seen at each of the frequencies shown in the maps. Some of the bursts at different frequencies are located close to each other.  
The detected bursts are usually unresolved within the instrumental resolution (Table \ref{Tab:psf}).
The strongest of the observed bursts are at $\sim$50 $\sigma_{map}$, where $\sigma_{map}$ is the RMS of the residual map away from the Sun. 
The 1-$\sigma$ fractional change in time for the maximum $T_B$ pixel from the visibility subtracted image lies between 2.13\% to 5.8\%. Such level of the variation is consistent with the temporal variations in the quiet Sun \citep{Oberoi2017,Sharma2018}.


\subsubsection{Noise statistics}
\begin{figure}
    \centering
    \resizebox{70mm}{!}{
    \includegraphics{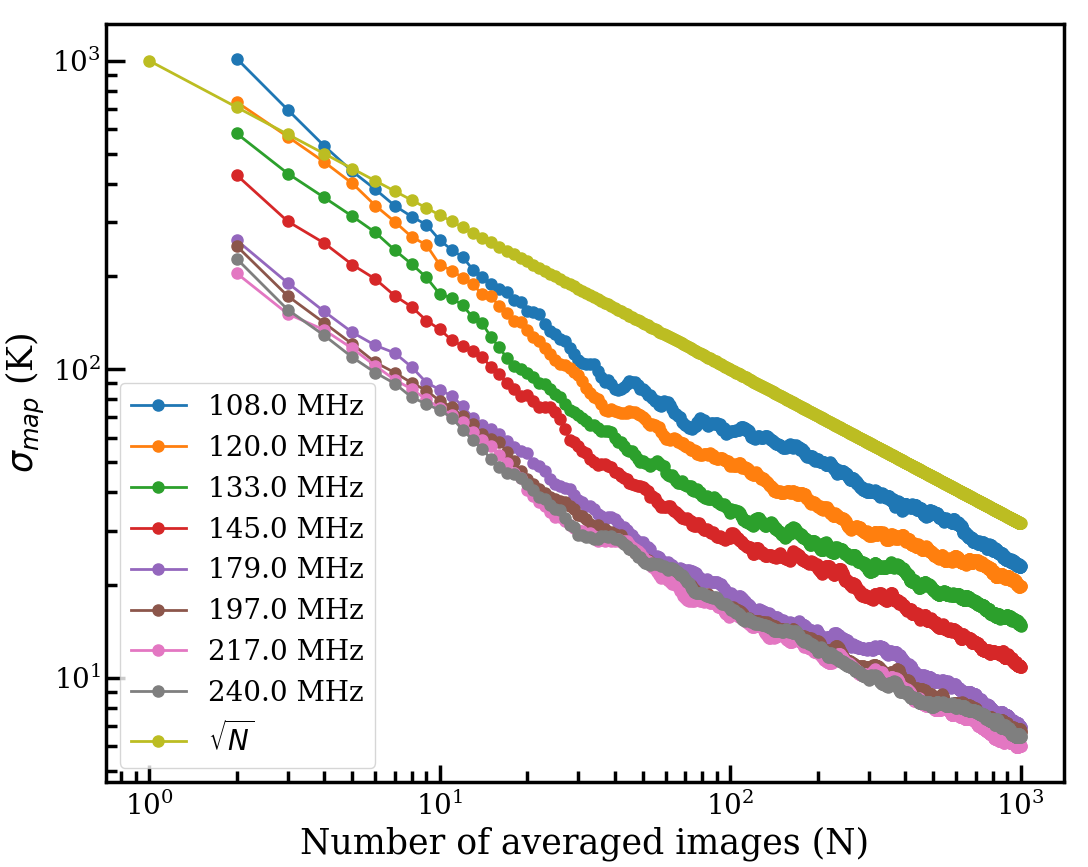}} 
    \caption{ $\sigma_{map}$ in the time-averaged image against the number of images averaged. The yellow line is proportional to $1/\sqrt{N}$, where N is number of averaged images. }
    \label{Fig:noise}
\end{figure}

The noise in the radio images is expected to be inversely proportional to the $\sqrt{\Delta \nu\ \Delta t}$, where $\Delta \nu$ and $\Delta t$ are frequency and time integration used for the image.
This holds true for the residual images as well.
Ideally, for well calibrated radio images, the noise in different images is independent and uncorrelated.
We explicitly verify if this is true for our data.
As $\Delta \nu$ for all of the images used here is the same, one expects as the observed noise in the image to reduce as $1/\sqrt{N}$, where $N$ is the number of images averaged.
The results are shown in Fig. \ref{Fig:noise} for each of the eight frequencies.
The reduction in noise is indeed found to closely follow a $1/\sqrt{N}$ trend, for each of the frequencies.
In addition, we notice a tendency for noise to decreases with increasing frequency.


\subsubsection{Self-noise}

``Self-noise'', also known as  ``source noise'', is the noise arising due to the stochastic nature of the flux density of the source itself.
In a typical non-solar radio image, the observed noise is dominated by instrumental contribution and the source noise is negligible.
In the limit where the source strength dominates other sources of noise present, the observed noise becomes independent of the sensitivity of the instrument and depends solely on source strength, 
i.e. $N_{self} = S/\sqrt{\Delta \nu\ \Delta t}$, where $S$ is the source strength and $N_{self}$ the source noise   \citep{Kulkarni1989, Morgan2021}. $S$ and $N_{self}$ can be expressed in either flux density or temperature units.
As emphasized by \citet{Kulkarni1989} and \citet{Morgan2021}, unlike thermal noise, source noise is not independent across baselines. 
Hence for radio interferometric images with a large number of elements, the source noise can become significant even when the noise on individual baselines is larger than the source noise.

The flux density of the Sun is comparable to the System Equivalent Flux Density (SEFD) of the MWA \citep{Oberoi2017}.
This makes it important to consider the contribution of the source noise in the present context, especially as we study weak radio bursts.
The expression given above, however, is valid only for an unresolved source, for which the same flux density is seen by each baseline.
It does not apply for the Sun, where the observed flux density falls by multiple orders of magnitude across the baseline lengths of the array (Fig. \ref{Fig:vis_hist}, panel C).
It is, however, hard to obtain a more realistic estimate of self noise for extended sources with complex structure, like the Sun, and we tabulate the numbers from the above expression in temperature units in Table \ref{tab:1}, as a useful upper limit for self noise.

Assuming Gaussian statistics, the number of spurious bursts within 30 min is $M_s=\eta M$, where $\eta$ and $M$ are the {\color{black} fraction of points beyond} 5-$\sigma$ and total data points in time respectively. Here, $\eta=6\times 10^{-5}$ and $M=3600$ for a given frequency band. Therefore, we would expect $\approx$ 0.2 bursts above 5-$\sigma_{map}$ levels in a 30 min interval. 
The number of bursts observed (described in Sec. \ref{sec:six_s}) is found to be well beyond what is expected from self-noise considerations (Fig. \ref{Fig:number} (A)).

\subsection{Identifying nonthermal features in residual images} \label{sec:six_s}


A signal-to-noise threshold needs to be imposed to distinguish real nonthermal features in the maps from noise fluctuations.
It could be imposed either in the image domain, based on the map RMS ($\sigma_{map}$), or in the time domain based on the RMS of light curve ($\sigma$) observed for a given pixel on the Sun. We introduce the following criteria for reliable detection of the nonthermal bursts.
\begin{enumerate}
    \item In each image, we select pixels $>$5$\sigma_{map}$. 
    \item  In time, we impose a threshold of 6$\sigma$ for a feature in the residual image to be regarded as a reliable detection. This was found to be the optimal choice which rejected the  few spurious features appearing outside the solar disk and retained the bursts within the disk.
    \item A further constraint requiring the size of bursts to be $\geq$ that of the PSF was also applied.
    As the final radio images images use a Gaussian restoring beam corresponding to the size of the PSF, any features smaller than the PSF are unreliable and were not considered in the downstream analysis.
\end{enumerate}
 
\begin{figure*}
    \centering
    \begin{tabular}{c}
    \resizebox{120mm}{!}{
         \includegraphics{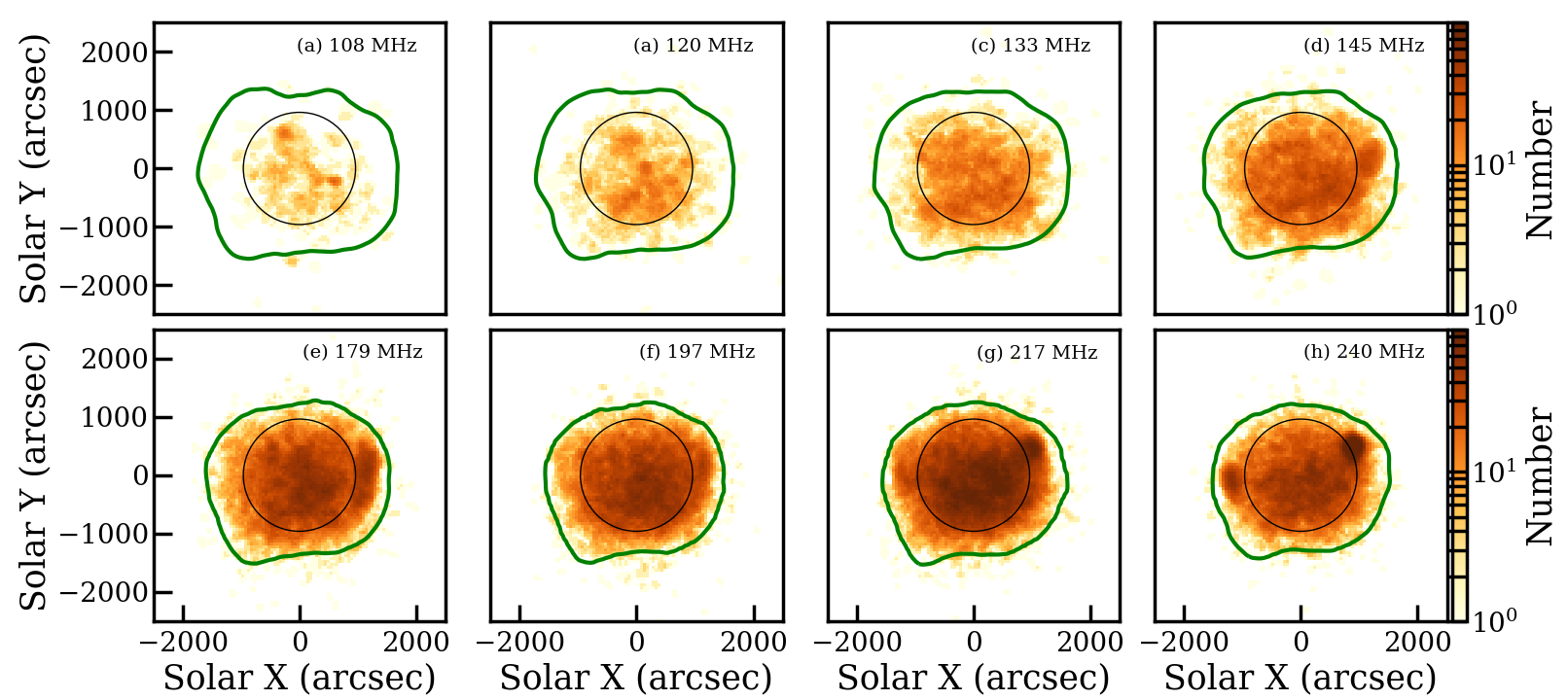}}\\
         (A) Number of detected bursts\\
        \resizebox{120mm}{!}{
         \includegraphics{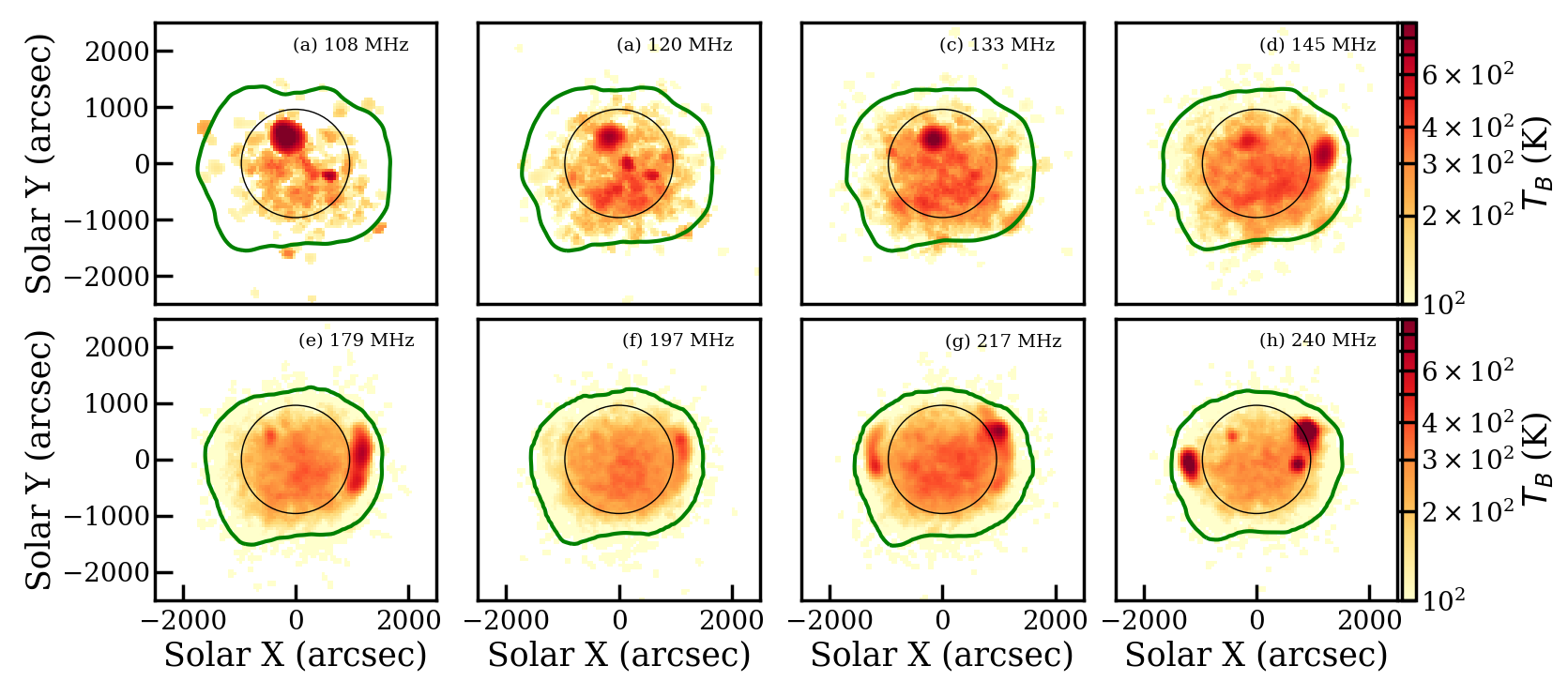}}\\
         (B) 1-$\sigma$ temporal variation of the detected bursts\\
    \end{tabular}
    \caption{Panel A: The map of total number of detected bursts at each of the frequency bands. 
    Panel B: 1$\sigma$ temporal variation map of the 6$\sigma$ and above features. The green contour in each panel marks the 5$\sigma_{map}$ boundary of radio solar maps, where $\sigma_{map}$ is the RMS in the off-sun region of the radio map. The black circle shows the optical disk position in each panel. Note that all colorbars are logarithmic.
    }
    \label{Fig:number}
\end{figure*}

The number of non-contiguous instances for which the emission from a given pixel exceeds the 6$\sigma$ is regarded as the number of reliably detected bursts from that location and is shown in Fig. \ref{Fig:number} (A) for all eight frequencies. 
To avoid apparent variations due to change in the instrumental resolution across our 108--240 MHz spectral range, the number of bursts in each pixel is  scaled inversely by number of pixels per beam for each frequency. Fig. \ref{Fig:number} (B) shows the temporal $T_B$ RMS map computed at each pixel in the residual maps.
At 145 MHz and higher, similar levels of RMS are seen over large parts of the Sun, though there are also some compact regions with higher RMS.

\begin{figure}
    \centering
    \resizebox{70mm}{!}{
\includegraphics[trim={0.0cm 0cm 0.0cm 0.0cm},clip,scale=0.3]{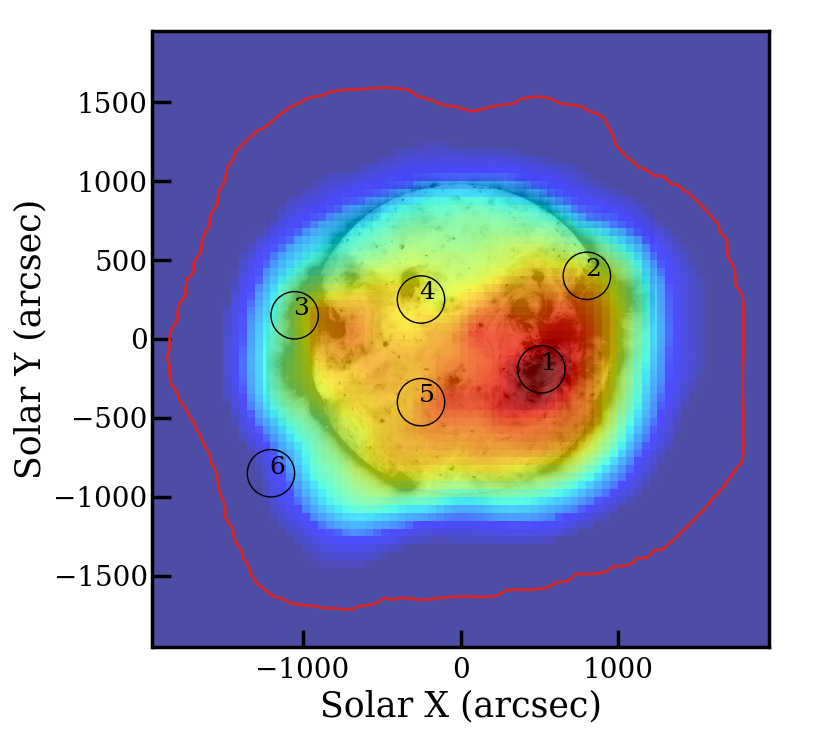}}
    \caption{ AIA 193 \AA\ image overlaid with 240 MHz radio map with the location of the six interesting regions. The red contour shows the 5-$\sigma_{map}$ boundary of 108 MHz radio maps, where $\sigma_{map}$ is the RMS in the off-sun region of the radio map. All regions are in within the solar radio disk.}
    \label{Fig:region}
\end{figure}

For a quantitative analysis across the disk, we select six regions (labeled 1 through 6) on the solar disk.
An example 240 MHz image is shown in Fig. \ref{Fig:region} and is used to identify these specific regions, each about the size of the point-spread-function.
Regions 1, 2 and  3 are in the vicinity of active regions. 
Region 1 corresponds to the brightest compact region on the radio images at high frequencies.
Regions 2 and 3 are located on the western and eastern limbs respectively.
Region 4 comes from an emerging active region, region 5 is a representative area from the quiet Sun, while region 6 is located off the solar limb.
The $T_B$ light curves from these regions are shown in panel A of Fig. \ref{Fig:ds_regions} for a few chosen frequencies.
The y-range varies across sub-panels and it is evident that bursts from regions 1, 2 and 3 are much brighter as compared to those from other regions. 
Table \ref{tab:1} lists the average $T_B$ for all bursts which meet the 6$\sigma$ threshold, along with the total number of such bursts for these six regions. Occasionally, we detect a few features below -6$\sigma$ in regions 1 through 4, however, their occurrence is low and does not significantly impact the averaged residual $T_B$ and number maps (Fig. \ref{Fig:number}). 

\begin{figure*}
    \centering
    \resizebox{120mm}{!}{
    \includegraphics{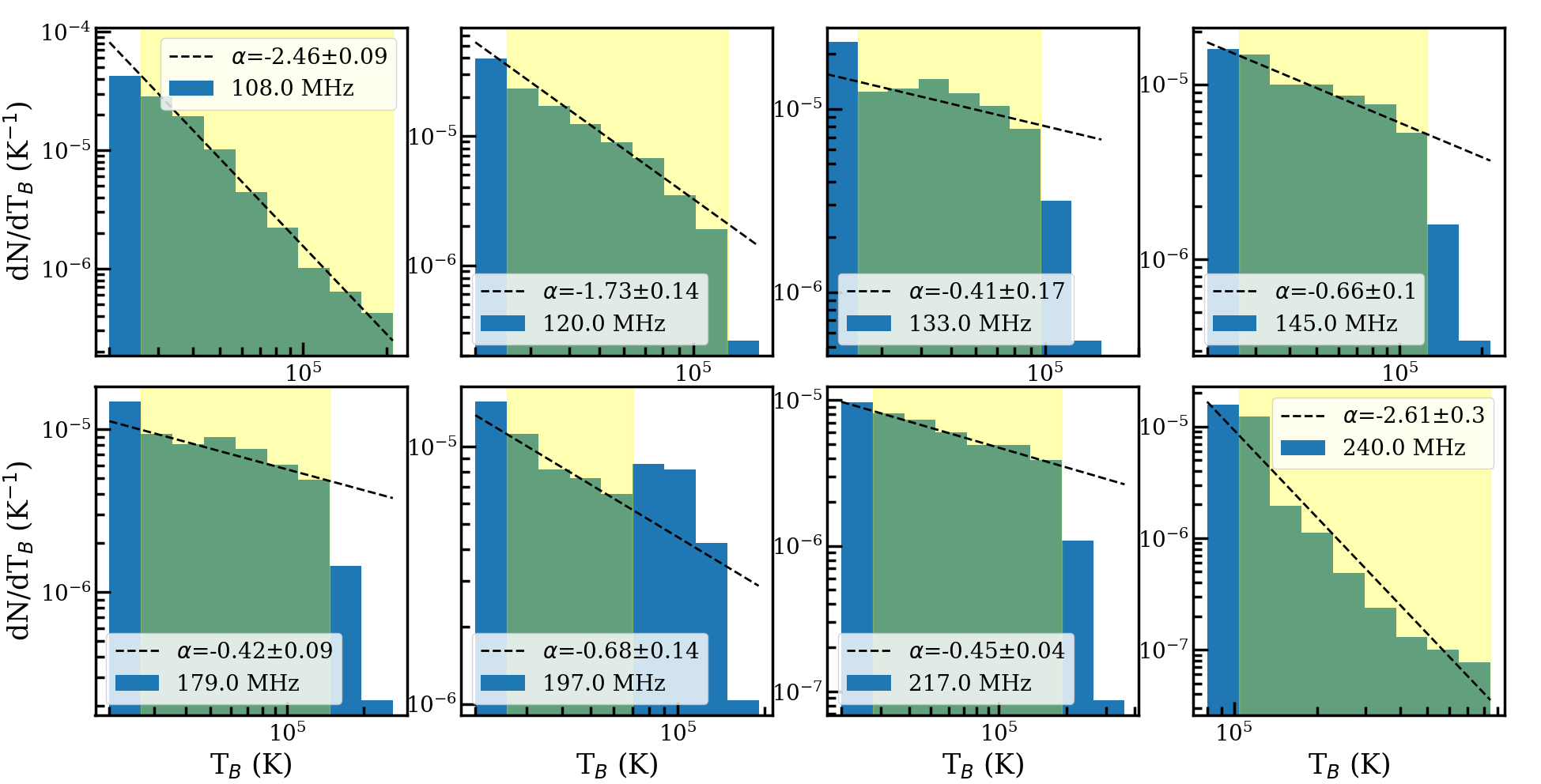}}
    \caption{$T_B$ distribution of all the reliably detected events at all frequencies along with a log-log fit. The yellow shaded region marks the $T_B$ range used to perform the power-law fit.}
    \label{Fig:Tb_dist}
\end{figure*}

A larger number of bursts are seen at the higher frequencies. 
The number of bursts significantly deceases at lower frequencies, however the time-averaged $T_B$ remains similar to that at higher frequencies -- $\sim$ few kK. Figure \ref{Fig:Tb_dist} plots the distribution of the $T_B$ for all of the detected bursts at each frequency.
 It is evident from Fig. \ref{Fig:number} that while much of the solar disk shows the presence of bursts, there also exist compact regions which show larger numbers of bursts. Such a region is especially prominent at 240 MHz, and is also found to be the site of comparatively brighter emission features (Fig. \ref{Fig:Tb_map}) and gives rise to the prominent bright tail in the histogram of $T_B$ distribution in Fig. \ref{Fig:Tb_dist}.
The power law index for the brightness temperatures for various frequencies varies widely between -0.42 and -2.6. 
 While these power law indices are provided, the comparatively narrow ranges over which they have been fit and the wide variations between the observed values suggest that we should not read much into them. 
These variations likely arise from multiple reasons. One important reason is the intrinsic sensitivity limitations of the data for detecting weak features, and the statistical fluctuations dominating the detection of strong features owing to the limited span of data analysed. In addition, these histograms combine the data from all of solar radio disk and hence encompass regions with substantially differing characteristics (Figs. \ref{Fig:region} and \ref{Fig:ds_regions} (A) and (B)). 

A few interesting events are also indicated in Figure \ref{Fig:ds_regions} (C) and will be discussed in Sec. \ref{sec:discussion}. 

 It is interesting to note that the $T_B$ associated with these features lie in the range of few time $10^4$--$10^5$K, even lower than the $10^6$K of the background coronal plasma. 
This is explained by the fact that these emissions are expected to arise from a compact region and that due to scatter broadening during coronal propagation their apparent observed size is much larger than their intrinsic size \citep[e.g.][]{Kontar2017, Sharma2020} and perhaps also instrumental resolution. The observed $T_B$ is hence highly diluted and does not represent the intrinsic brightness temperature of the emission. In addition to dilution along the angular size axis, as these sources are unresolved in time \citep{Mondal2020b} and expected to be narrowband, similar arguments for dilution of $T_B$ along the time and frequency axis also apply. Together, these effects explain the low observed $T_B$ of these features.

\subsection{Duration of bursts}

\begin{figure*}
\begin{tabular}{c}
\resizebox{180mm}{!}{
\includegraphics[trim={0.0cm 0cm 0.0cm 0.0cm},clip,scale=0.3]{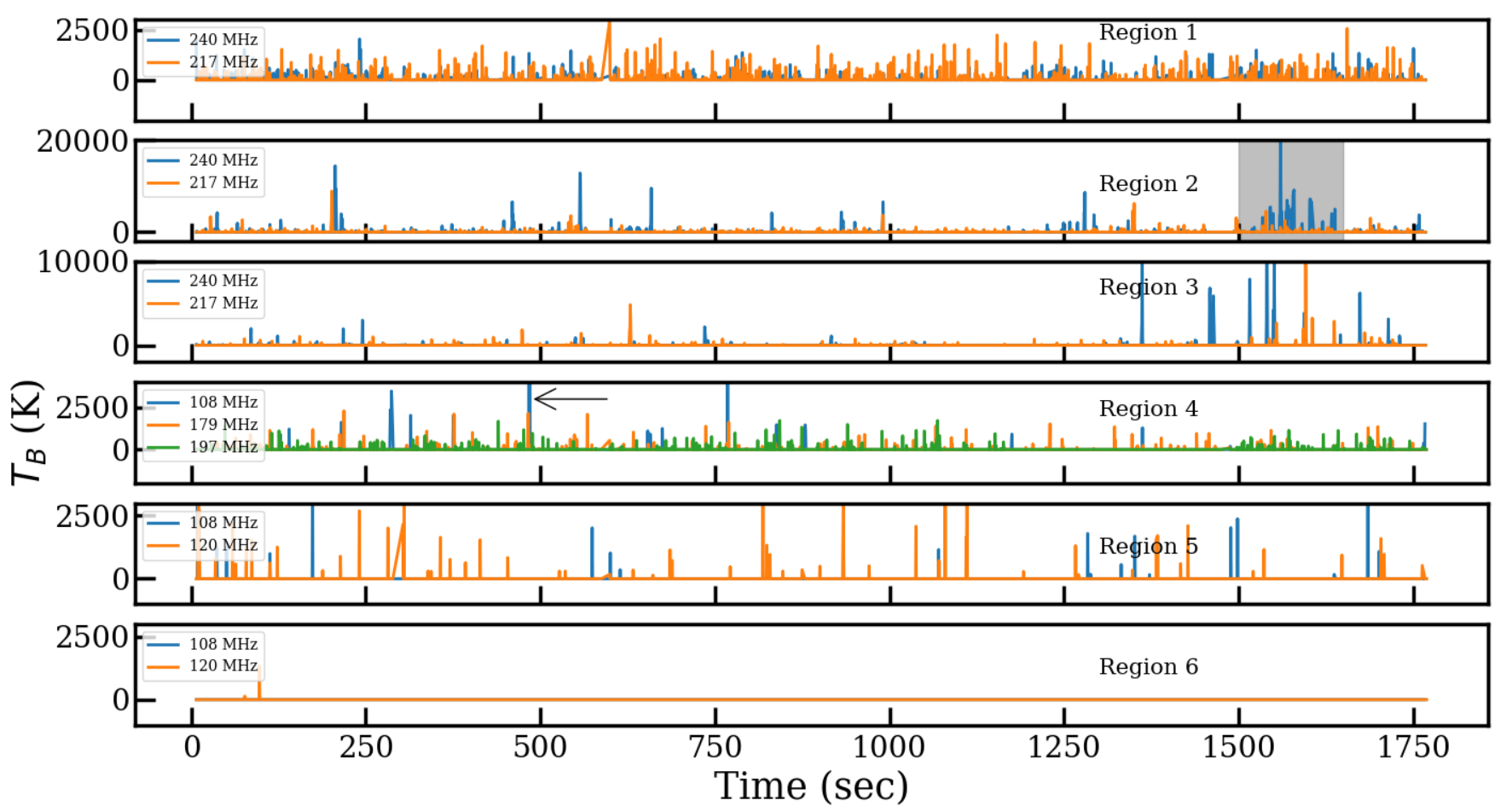}}\\
(A) $T_B$ time-series of various regions
\end{tabular}
\begin{center}
\begin{tabular}{cc}
\resizebox{110mm}{!}{
\includegraphics[trim={0.0cm 0cm 0.0cm 0.0cm},clip,scale=0.3]{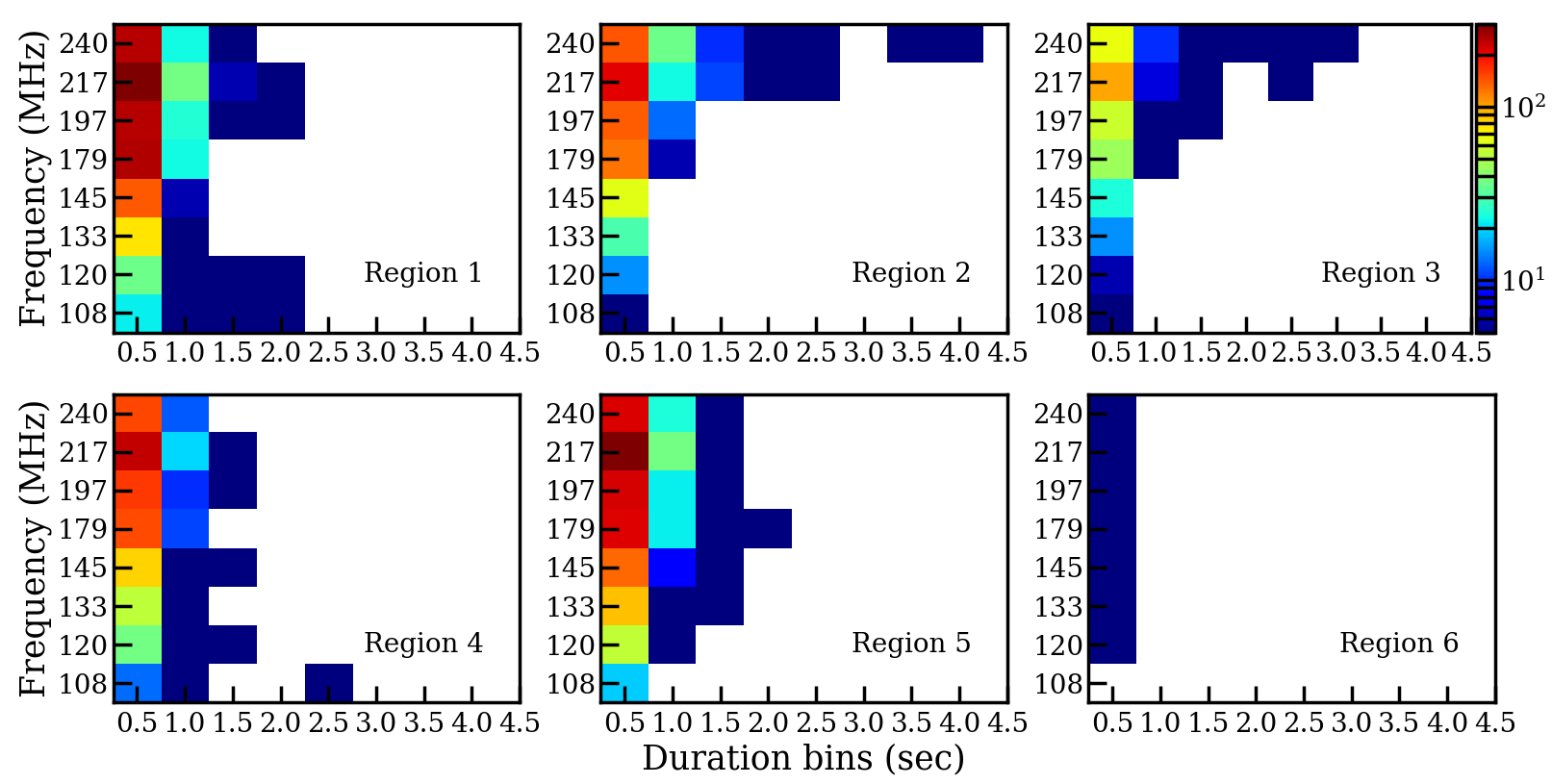}}
&
\resizebox{65mm}{!}{
\includegraphics[trim={0.0cm 0cm 0.0cm 0.0cm},clip,scale=0.3]{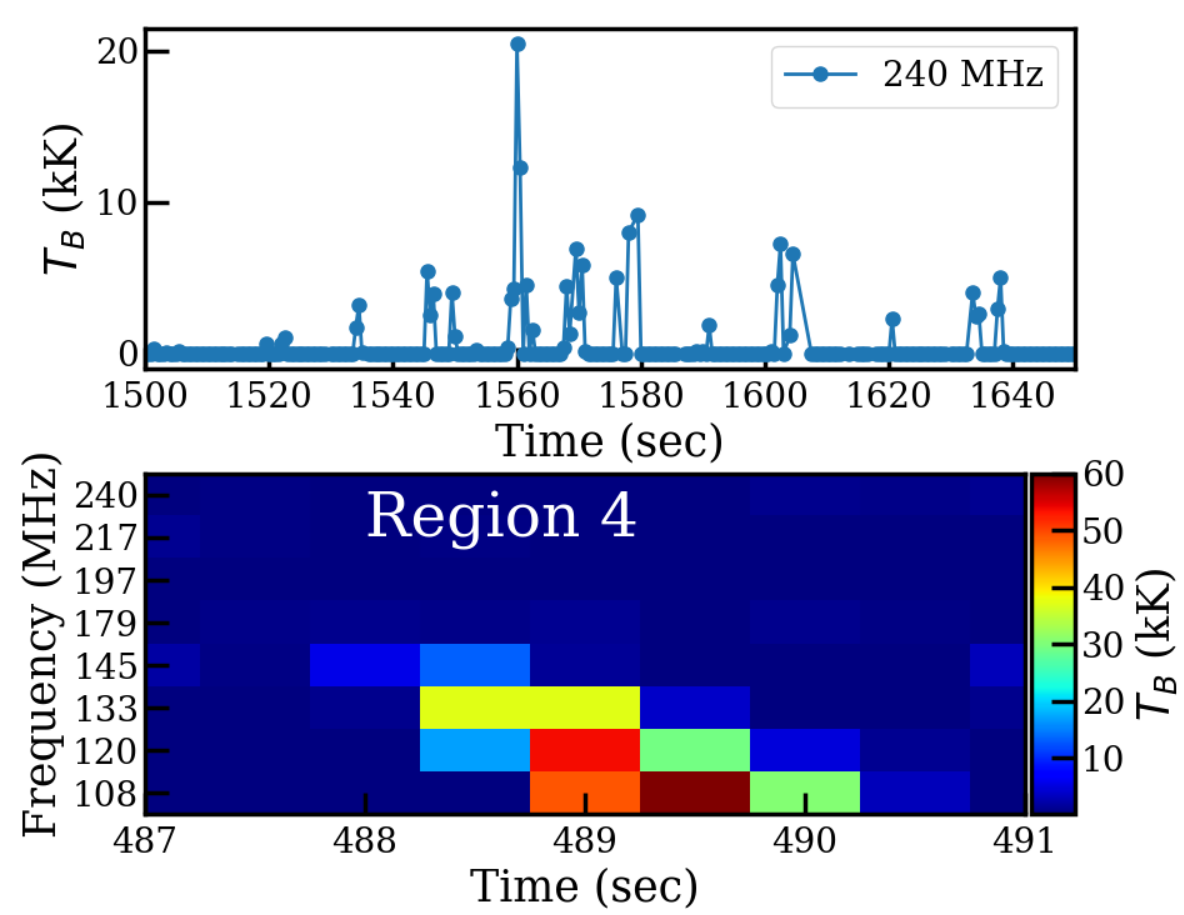}} 
 \\
(B) 2-D histogram in frequency and duration.   & (C) 240 MHz quasi-periodic bursts and a drift burst
\end{tabular}
\end{center}
\caption{Panel (A) Time series of the $T_B$ of the impulsive events for various regions in Fig.\ref{Fig:region}. The gray shaded region marks the times of quasi-periodic bursts mentioned in Section \ref{sec:discussion}. The arrow in region 3 shows the drift burst location in time. Panel (B) 2-D histogram of the number of impulsive emission for all the regions in frequency and duration.  Panel (C) Top panel shows the zoomed-in timeseries of quasi-periodic bursts at 240 MHz. Bottom panel shows the drift burst in the region 3.
\label{Fig:ds_regions}}
\end{figure*}

The method employed here can efficiently identify impulsive emissions on time scales significantly smaller than 15 s.
We regard the number of contiguous time slices for which the observed residual intensity meets the 6$\sigma$ threshold as the duration of an individual burst.
The durations thus estimated for the regions identified in the previous subsection are shown as a 2D histogram in Fig. \ref{Fig:ds_regions} (B).
The vast majority of events are unresolved by the 0.5 s resolution of the data.
Even at the higher frequencies, where some longer duration events are observed, the duration of  the longest bursts does not exceed 4 s and the number of observed events always drops very steeply with increasing duration.
This suggests that there is little possibility of missing longer duration events due to the use of the 15 s running median window.
Absence of bursts of durations longer than a few seconds has been seen before.
\citet{Suresh2017} found in their non-imaging studies of weak type I like emissions that the distribution of burst durations peaked in the range 1-2 s and fell by two orders of magnitude by about 5 s. In an imaging study of much weaker quiet Sun emissions, \citet{Mondal2020b} found the peak lie at instrumental resolution of 0.5 s and dropped by two orders of magnitude by 5 s, similar to the present case.
The short timescales of these emissions are likely related to the longevity of electron beams giving rise to them -- the weaker the electron beams, the shorter their lifespan.

\subsection{Time-averaged maps and EUV counterparts}

Making time-averaged maps allows us to find regions of persistent emission. We make two time-averaged maps from residual maps for each frequency -- one averaging over the entire observing duration, while the other one is averaged over only the reliable bursts identified in Sec. \ref{sec:six_s}.
These maps are
shown in Figs. \ref{Fig:averaged_map} and \ref{Fig:Tb_map} respectively.

\subsubsection{Total time-averaged map}

\begin{figure*}
    \centering
    \begin{tabular}{c}
    \resizebox{120mm}{!}{
         \includegraphics{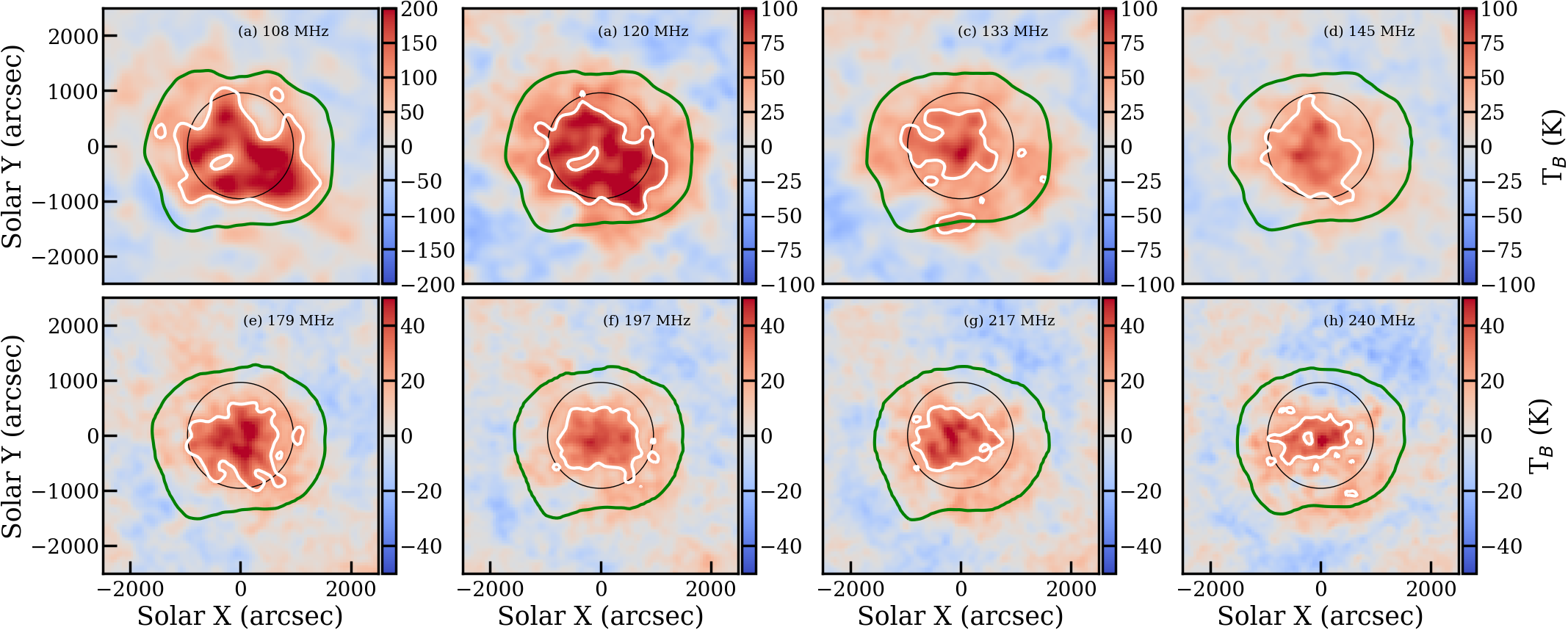}}\\
          (A) Total time-averaged map \\ 
    \resizebox{120mm}{!}{
          \includegraphics{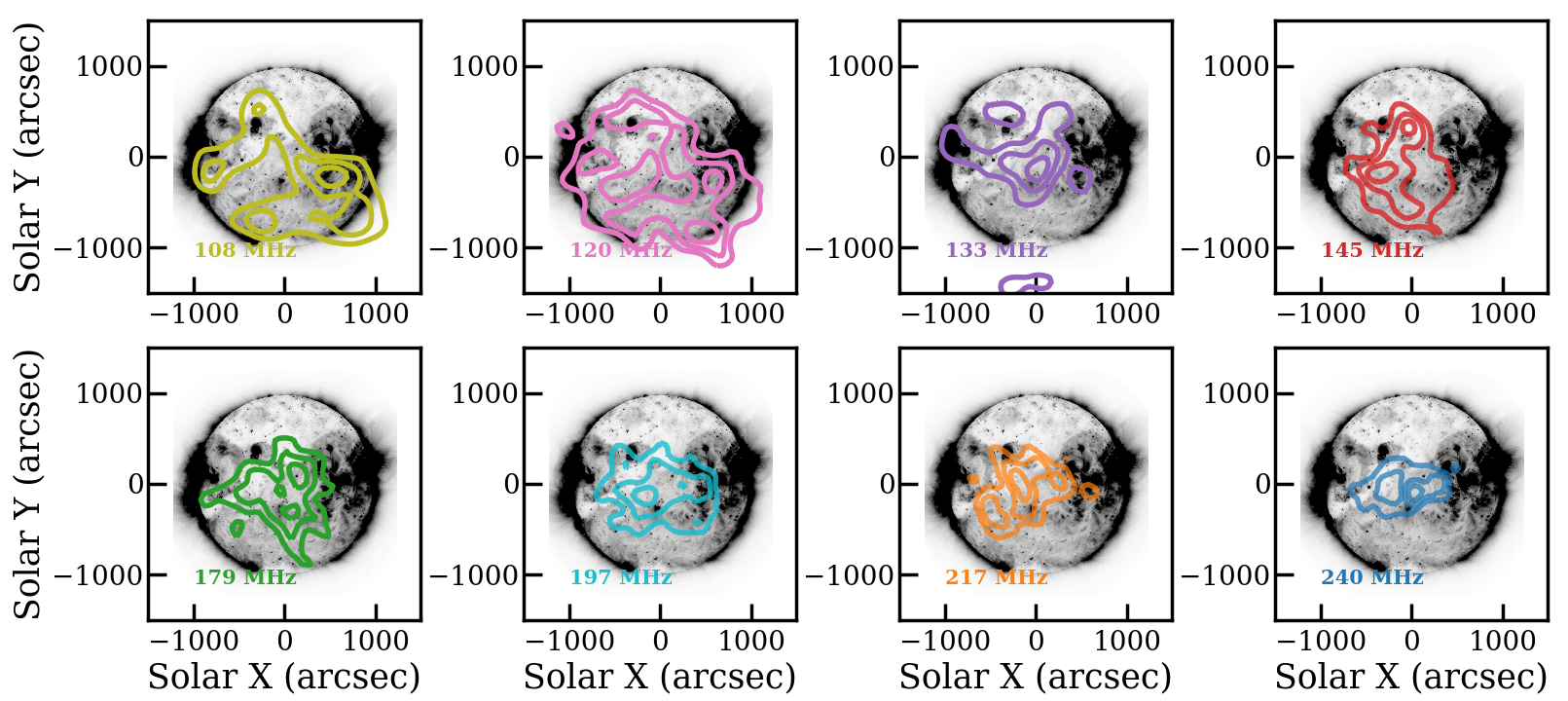}}\\
         (B) AIA 193 \AA\ EUV map overlaid with radio contours
   \end{tabular}
    \caption{Panel A:  Time-averaged residual $T_B$ maps for various frequency bands after removing $6-\sigma$ and above time stamps. The white contour in each panel shows the 5$\sigma_{map}$ contour, where $\sigma_{map}$ is the RMS computed in the off-solar disk region in the residual solar maps. The green contour in each panel marks the 5$\sigma_{map}$ boundary of radio solar maps. The black circle marks the edge of the solar optical disk. Panel B: The contours of the time-averaged residual maps overlaid on the AIA 193 \AA\ image. The contours are plotted at 0.6, 0.75 and 0.9 times the maximum. }
    \label{Fig:averaged_map}
\end{figure*}

Figure \ref{Fig:averaged_map} shows the total time-averaged map after removing $6-\sigma$ and above time stamps for all eight frequency bands. 
At all frequencies, the observed average residual emission in the solar region is overwhelmingly positive. 
At 108 MHz and 120 MHz, the bulk of the solar disk is covered by significant emission, while at higher frequencies the area covered by this emission is considerably smaller.
In all cases the central region of the Sun is covered by this emission.
The observed strength of the emission is also much higher at 108 MHz and 120 MHz as compared to higher frequencies.

Panel B of Fig. \ref{Fig:averaged_map} compares the radio maps with EUV 193 \AA\  maps, which are sensitive to MK coronal plasma. 
We do not find correspondence between them and bright $T_B$ regions 1 to 4 seen in Fig. \ref{Fig:region} in the total averaged radio maps (Fig. \ref{Fig:averaged_map}). Distinct dip in the emission can also be seen at the coronal hole region, especially the coronal hole section which crosses the equator (Figure \ref{Fig:averaged_map} (A) and (B) 108 MHz) at (-700",-200"). 


\subsubsection{Burst's time-averaged map}

\begin{figure*}
    \centering
    \begin{tabular}{c}
        \resizebox{120mm}{!}{
         \includegraphics{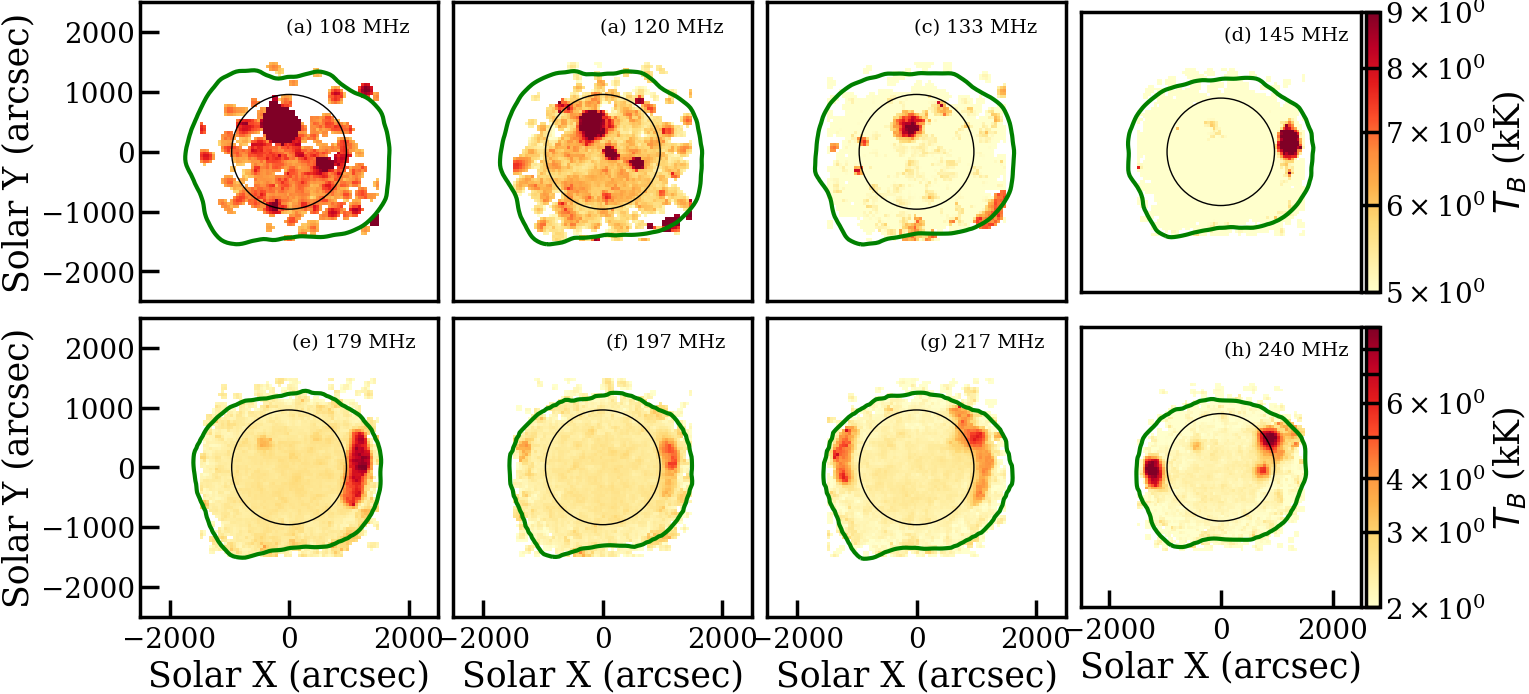}}\\
         (A) $T_B$ of the detected bursts\\
    \resizebox{110mm}{!}{
          \includegraphics{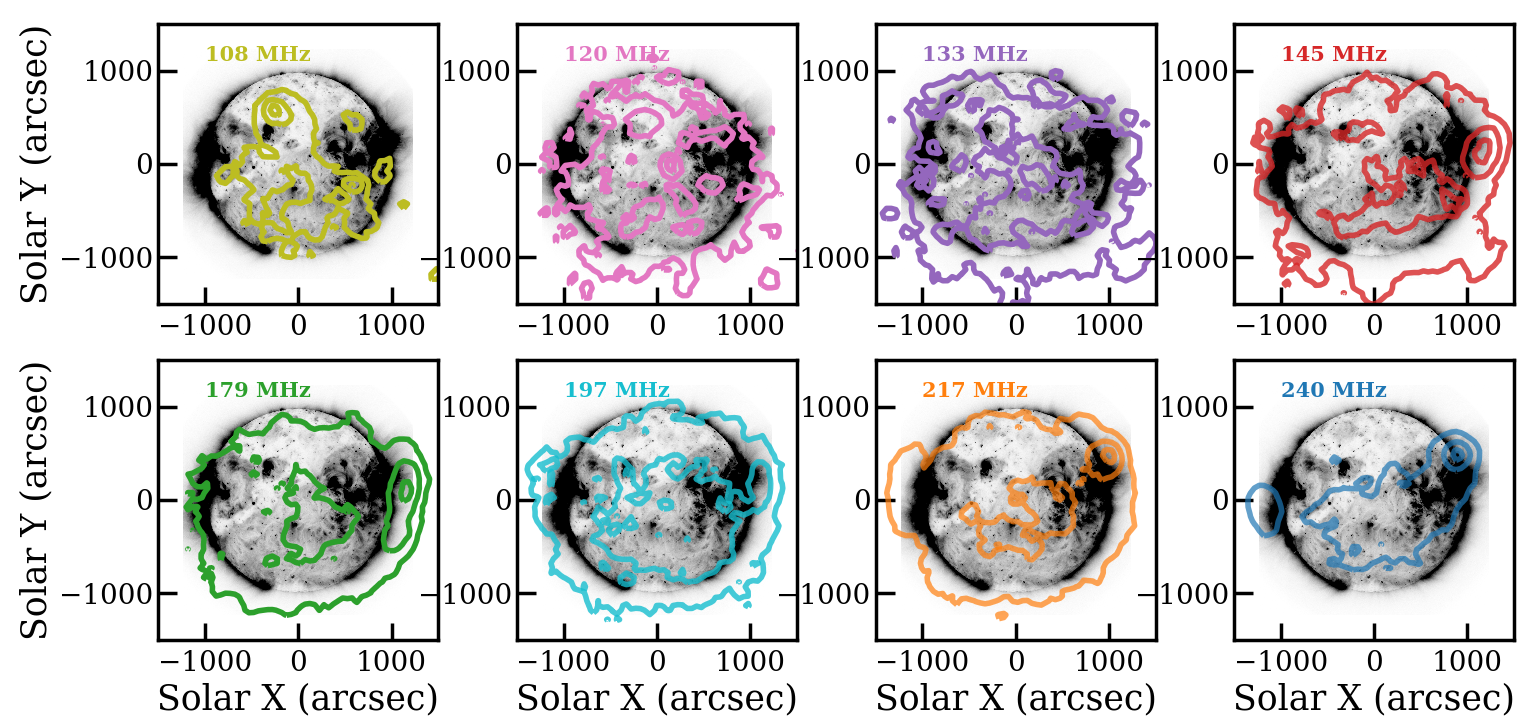}}\\
           (B) $T_B$ of the detected bursts overlaid with AIA 193 \AA \ image \\
    \end{tabular}
    \caption{
    Panel A: Time-averaged $T_B$ of the 6$\sigma$ and above bursts.
    Panel B: The contours of the maps of $T_B$ of the 6$\sigma$ and above bursts overlaid on the AIA 193 \AA\ image. The contour plotted are 0.1, 0.5 and 0.9 times w.r.t map's maximum. The green contour in each panel marks the 5-$\sigma_{map}$ boundary of radio solar maps, where $\sigma_{map}$ is the RMS in the off-sun region of the radio map. The black circle shows the optical disk position in each panel. Note that all colorbars are logarithmic.
    }
    \label{Fig:Tb_map}
\end{figure*}

Panel A of Fig. \ref{Fig:Tb_map} shows the time-averaged map of the bursts obtained by following the criteria explained in Sec. \ref{sec:six_s}. The panel B shows the radio contours overlaid on 193\ \AA \ EUV map.
We note the bursts occur almost uniformly across the solar disk, with some brighter patches. Those bright patches are seen from 145 MHz to 240 MHz on the western limb near region 1, also on the eastern limb at 217 MHz and 240 MHz. At 108 MHz and 120 MHz, some isolated patches are seen near region 3 and in the southern hemisphere. The occurrence of brighter patches at a similar location across all frequencies suggest a continuity in the emission across a large range of coronal heights. 
We note that the bright spots at 240 MHz are more prominent in the $T_B$ map as compared to maps showing the number of bursts (Fig. \ref{Fig:number} (A)), suggesting the presence of large fraction of fainter bursts spread over solar disk. 
Figure \ref{Fig:Tb_map} (B) shows the radio contours overlaid on a 193 \AA\ EUV map. The brighter patches can be clearly seen at 240 MHz, 217 MHz and 108 MHz corresponding to active regions on the western and eastern limbs and an on-disk emerging active region. Figure \ref{Fig:number} (B) shows the 1$\sigma$ level of temporal fluctuation of the $T_B$ of the bursts. Among all, this map shows the bright spots most clearly. 



\begin{table*}
  \centering
  \begin{tabular}{|p{1.2cm}|p{1.2cm}|p{1.2cm}|p{1.5cm}|c|c|c|c|c|c|c|c|c|c|c|c|}
    \hline
    \multirow{1}{*}{Frequency (MHz)}&  \multirow{1}{*}{Self Noise (kK)} & 
    \multirow{1}{*}{f$_{B,max}$ (\%)} & 
    \multirow{1}{*}{f$_{B,S,max}$(\%)} & \multicolumn{2}{c|}{Region 1} & \multicolumn{2}{c|}{Region 2} & \multicolumn{2}{c|}{Region 3} & \multicolumn{2}{c|}{Region 4} & \multicolumn{2}{c|}{Region 5} & \multicolumn{2}{c|}{Region 6}\\
    \cline{5-16}
    & & & & $T_B (kK)$ & N$_{bt}$ &  $T_B (kK)$ & N$_{bt}$ &  $T_B (kK)$ & N$_{bt}$ & $T_B (kK)$ &N$_{bt}$& $T_B (kK)$ & N$_{bt}$ & $T_B (kK)$ & N$_{bt}$\\
    \hline

108.0  & 0.26 & 2.28 & 4.15 & 4.73  $\pm$ 0.2 & 11 & 8.08  $\pm$ 0.1 & 1 & 0 & 0 & 107.53  $\pm$ 0.1 & 1 & 33.12  $\pm$ 0.17 & 1 & 0.0  $\pm$ 0.0 & 1 \\ \hline
120.0  & 0.3 & 2.33 & 5.03 & 8.09  $\pm$ 0.21 & 8 & 5.33  $\pm$ 0.1 & 3 & 3.73  $\pm$ 0.09 & 3 & 19.06  $\pm$ 0.1 & 5 & 5.39  $\pm$ 0.22 & 14 & 0.74  $\pm$ 0.02 & 2 \\ \hline
133.0  & 0.32 & 2.16 & 4.09 & 3.9  $\pm$ 0.23 & 22 & 4.95  $\pm$ 0.12 & 5 & 3.99  $\pm$ 0.04 & 2 & 6.67  $\pm$ 0.1 & 15 & 5.73  $\pm$ 0.2 & 16 & 1.03  $\pm$ 0.01 & 1 \\ \hline
145.0  & 0.32 & 2.13 & 2.97 & 4.06  $\pm$ 0.24 & 31 & 4.86  $\pm$ 0.14 & 11 & 5.53  $\pm$ 0.07 & 3 & 9.77  $\pm$ 0.18 & 10 & 4.02  $\pm$ 0.23 & 29 & 0.15  $\pm$ 0.01 & 9 \\ \hline
179.0  & 0.34 & 2.32 & 5.8 & 2.71  $\pm$ 0.21 & 56 & 3.62  $\pm$ 0.12 & 17 & 1.52  $\pm$ 0.06 & 12 & 1.94  $\pm$ 0.15 & 39 & 2.97  $\pm$ 0.17 & 43 & 0.11  $\pm$ 0.01 & 7 \\ \hline
197.0  & 0.4 & 2.3 & 2.19 & 2.79  $\pm$ 0.19 & 52 & 2.43  $\pm$ 0.13 & 29 & 2.09  $\pm$ 0.07 & 11 & 1.95  $\pm$ 0.11 & 39 & 2.84  $\pm$ 0.19 & 48 & 0.24  $\pm$ 0.01 & 4 \\ \hline
217.0  & 0.45 & 2.51 & 4.13 & 2.57  $\pm$ 0.23 & 83 & 3.74  $\pm$ 0.36 & 54 & 3.13  $\pm$ 0.3 & 29 & 1.94  $\pm$ 0.17 & 54 & 2.38  $\pm$ 0.19 & 69 & 0.02  $\pm$ 0.0 & 10 \\ \hline
240.0  & 0.47 & 2.45 & 4.51 & 2.12  $\pm$ 0.16 & 54 & 5.13  $\pm$ 0.91 & 85 & 6.15  $\pm$ 0.68 & 30 & 1.86  $\pm$ 0.09 & 30 & 2.79  $\pm$ 0.15 & 35 & 0.15  $\pm$ 0.01 & 6 \\ \hline

  \end{tabular}
  \caption{Various estimates for MWA frequencies. The second column from left lists estimates of self-noise. The third column f$_{B,max}$ is the percentage RMS change in $T_B$ for the image maximum pixel, defined as $f_{B,max} = \frac{\delta T_B}{T_{B1}}*100$, where $T_{B1}$ and $\delta T_B$ are the brightness temperature at the location of the maxima in the image and the 1$\sigma$ RMS at the same location respectively.
  Similarly, the fourth column shows the percentage change obtained from the visibility subtracted images, i.e. f$_{B,S,max} = \frac{T_{B,sub,max}}{T_{B1}}*100$, where $T_{B,sub,max}$ is the maximum $T_B$ in the visibility subtracted images. From fifth column onwards, the table lists the time-averaged $T_B$ and number of the impulsive features for various regions shown in Fig. \ref{Fig:region}.}
  \label{tab:1}
\end{table*}

\begin{table*}
    \centering
    \scalebox{0.7}{
    \begin{tabular}{|c|c|c|c|c|c|c|c|c|c|c|}
    \hline
         Frequency (MHz) & $n_e (\times 10^{8}$ cm$^{-3}$) & $B$ (G) & $T_e$ (MK) & $E_{b}$ ($\times 10^{25}$ ergs) & $E_{th}$ ($\times 10^{25}$ ergs) & Flux density (mSFU) & $W$ ($\times 10^{15}$ ergs) & $\gamma=\frac{W}{E_{nth}} (\times 10^{-5})$ & $T_{6s,B}$ (kK) & $E_{nth}$ ($\times 10^{20}$ ergs) \\ \hline \hline

108.0  & 1.43  &  2.7 & 1.6  &  7.48  &  0.82  &  0.49  &  0.1  &  0.38  &  8.47  &  0.26 \\ \hline
120.0  & 1.8  &  2.96 & 1.62  &  8.99  &  1.04  &  0.93  &  0.29  &  0.75  &  6.22  &  0.38 \\ \hline
133.0  & 2.21  &  3.18 & 1.59  &  6.92  &  0.84  &  1.82  &  0.62  &  1.93  &  5.01  &  0.32 \\ \hline
145.0  & 2.59  &  3.34 & 1.53  &  7.65  &  0.94  &  2.96  &  1.12  &  1.89  &  4.07  &  0.59 \\ \hline
179.0  & 4.06  &  3.69 & 1.16  &  4.67  &  0.56  &  4.96  &  2.07  &  3.33  &  2.67  &  0.62 \\ \hline
197.0  & 4.8  &  3.75 & 1.01  &  9.62  &  1.15  &  5.64  &  6.41  &  2.18  &  2.48  &  2.93 \\ \hline
217.0  & 6.71  &  3.8 & 0.76  &  4.94  &  0.61  &  9.74  &  44.86  &  2.84  &  2.43  &  15.77 \\ \hline
240.0  & 6.71  &  3.8 & 0.76  &  9.89  &  1.22  &  9.56  &  17.9  &  0.6  &  2.4  &  29.98 \\ \hline

    \end{tabular}}
    \caption{List of various coronal parameters (electron density, magnetic field strength and temperature) and derived magnetic field energy, thermal energies. An average flux density of the bursts, radiation energy, fraction of radiation energy and total energy are also listed for MWA frequencies.}
    \label{tab:2}
\end{table*}

\section{Burst energetics} \label{sec:energetics}

Narrowband and short-lived radio bursts at meterwaves are produced by plasma emission mechanisms. The travelling electron beams in the corona produce coherent radio bursts at the local plasma frequency due to beam-plasma instability \citep[e.g.][]{McLean1985, Dulk1985, Melrose2017}. 
However, the amount of energy radiated away as radio bursts is minuscule as compared to total bolometric energy of the bursts \citep{Bastian1998}. In this section, we compute the nonthermal particle energies, burst radiation energy, and total free magnetic and thermal energy available for the bursts for the date of observation. We obtain the physical state of the corona from the FORWARD model using
the coronal model from PSIMAS \footnote{https://www.predsci.com/corona/model$_-$desc.html} for our observing date \citep{Gibson2015,Gibson2016}. 
The same coronal model was used in \cite{Sharma2020}, which focused on investigating coronal propagation effects. 
In MAS models, the coronal physical parameters are computed in a manner designed to provide a self-consistent description of the coronal environment. This model is used to obtain the 3D distribution for the electron density ($n_e$), coronal temperature ($T_e$) and vector magnetic fields ($\vec{B}$) along latitude, longitude and height axes.

\begin{figure}
    \centering
    \begin{tabular}{c}
    \resizebox{70mm}{!}{
    \includegraphics{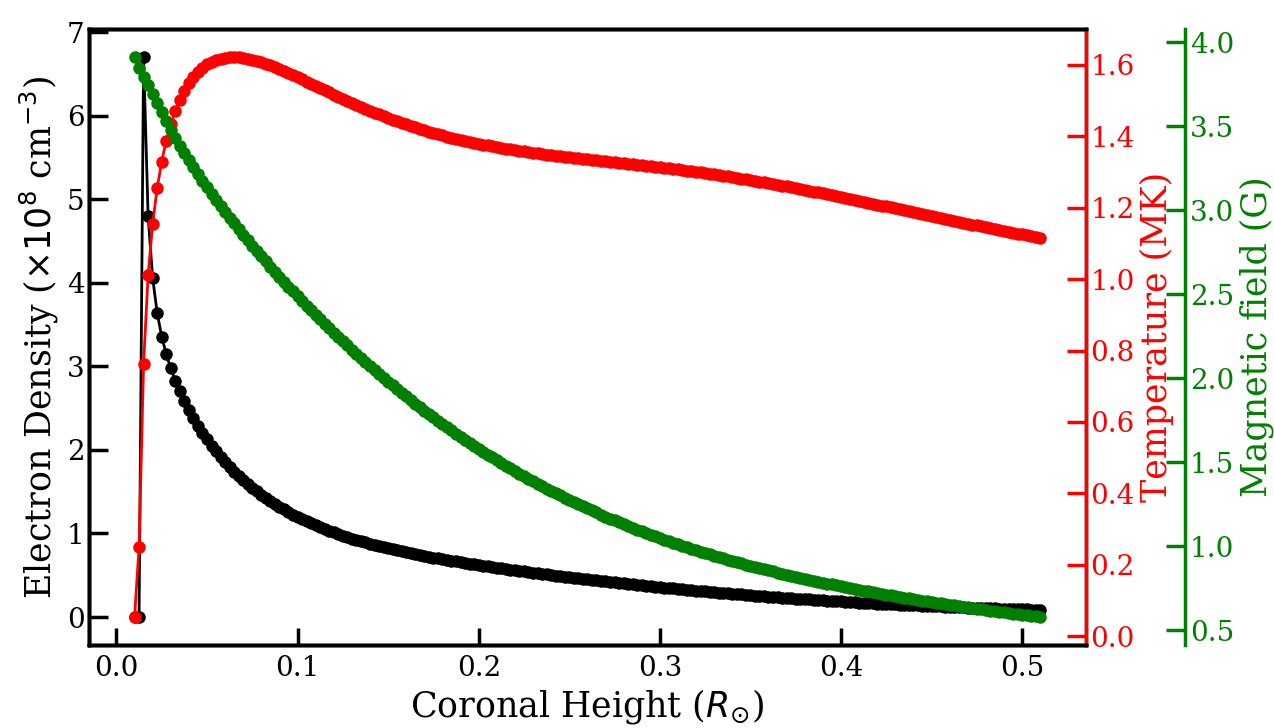}} \\
       (A) FORWARD parameters \\
    \resizebox{55mm}{!}{
    \includegraphics{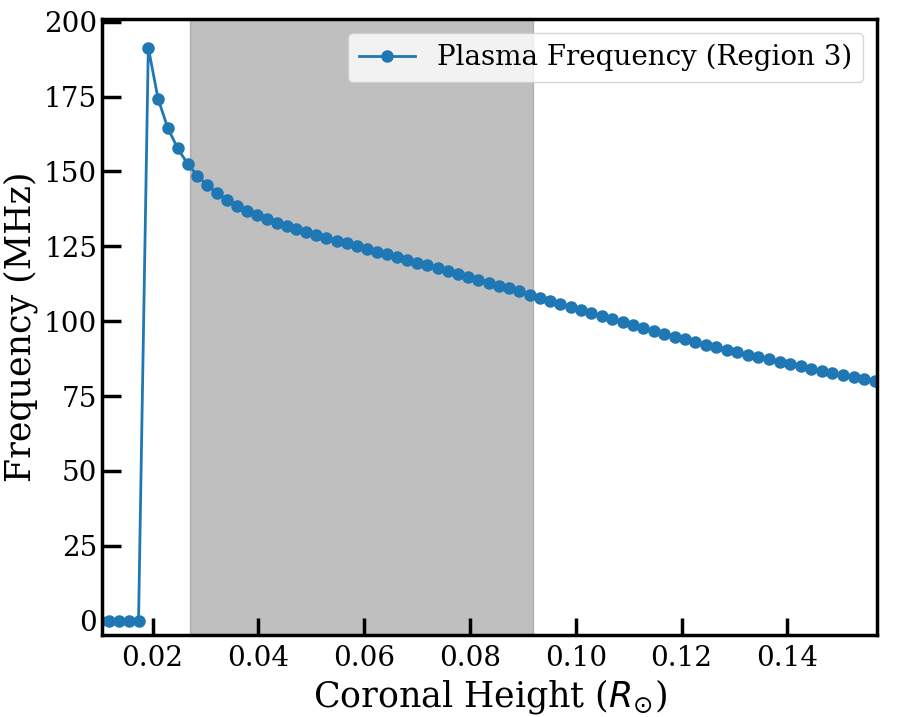}}  \\ (B) $\omega_p$ (Region 4)
    \end{tabular}
    \caption{ Panel A: Typical quiet Sun electron density (black), temperature (red) and magnetic field strength (green) model obtained from PSIMAS model of $3^{rd}$ December 2015. Panel B: For Region 4 the plasma frequency as a function of coronal height obtained from the FORWARD model. The coronal height is measured from the solar surface. The gray shaded area shows the coronal height range corresponding to the type-III burst (Fig. \ref{Fig:ds_regions} (C)). }
    \label{Fig:fwd}
\end{figure}
  
Using these 3D distributions, we construct an averaged 1D profile for $n_e$, $T_e$ and $|B|$ as a function of coronal height, which is shown in Fig. \ref{Fig:fwd} (A). We select the coronal heights for fundamental plasma frequencies corresponding for the observed frequencies and obtained coronal parameters $n_e$, $T_e$ and $|B|$ at MWA frequencies which are listed in Table \ref{tab:2}.
 
 \subsection{Nonthermal Flare Energy}

The calculation of the magnetic energy released to the nonthermal particles in the flare is nontrivial. 
However estimates of the nonthermal energy of the particles can be obtained under the isothermal, homogeneous and optically thick plasma assumption, by associating variation in $\Delta T_B \sim \Delta T_e e^{-\tau}$, where $\tau_0$ is the optical depth {\color{black} and $\Delta T_B$ represents in the approximate increase in temperature due to the release of magnetic energy}.

For this calculation, we assume an optical depth for quiet Sun for the meterwave frequencies based on \citep{Oberoi2017}. 
For each radio burst, we assume an associated population of {\color{black}supra}thermal electrons of density ($n_{nth}$) occupying a volume $(V)$ at a temperature increment corresponding $\Delta T_B$. For such an electron population, the nonthermal energy is given by $E_{nth} = (n_{nth} k_B \Delta T_B e^{\tau_0}) *A *l$, where $A$ and $l$ is the area and column depth of the radio source, and $V=A\times l$ \citep{Ramesh2010}.
Due to the absence of EUV counterparts, the area of the flaring region is difficult to accurately estimate. Therefore, we assume a typical flaring area region of 100 arcsec$^2$, i.e $A$ = 5$\times10^{17}$ cm$^{2}$
\citep[e.g.][]{Warmuth2020} and filling-factor of 1. From the previous non-imaging analysis, the bandwidths of the radio bursts during quiet time was found to be $\sim$ 5 MHz \citep{Suresh2017}. For the calculation of $l$, we assume differential radial heights corresponding to plasma frequencies for the narrowband burst of averaged bandwidth, i.e. 5 MHz. We get typical column depth as $3\times 10^8$ cm.

To estimate the nonthermal electron density, we assume an efficiency of $\beta$, i.e. the nonthermal density $n_{nth} \approx \beta n_{th}$. However, calculation of $\beta$ is nontrivial for quiet Sun regions. For flares in active regions, $\beta$ can be as large as $\sim 10^{-3}$ \citep{Sharma2020M}. If we regard the bursts as travelling electron beams exciting Langmuir wave modes, then from the simulation studies, $\beta \approx 10^{-5} - 10^{-7}$ \citep{Li2009,Ramesh2010,Volokitin2018}.
We use values $n_e$, $A$ and $l$ mentioned in the coronal model in Sec. \ref{sec:model} and a nominal value of $\beta \sim 10^{-5}$. Using nonthermal density and source volume, we get total electron number $\approx 1.5\times 10^{29}$. Table \ref{tab:2} lists the values of average $T_B$ for the detected bursts ($T_{6s,B}$), which lie in kilo-Kelvin temperature range. We note that the nonthermal energies lie in the range $\sim10^{20}-10^{21}$ ergs, i.e. in the sub-picoflare range. 
If we consider $\beta \sim 10^{-7}$, the nonthermal energies drop to lie in the  $\sim10^{18}-10^{19}$ ergs range. By taking the variation in $\beta$, we obtain a nonthermal energies lies between $\sim10^{18}-10^{21}$ ergs.

For sustained coronal heating, the distribution of flare energies must have a slope steeper than -2
\citep{Hudson1991}. In Fig. \ref{Fig:occurance} (B), we plot the distribution of nonthermal flare energies.
A linear fitting was done from $10^{20}$ to $10^{21}$ ergs energy range. We obtained a slope of -1.99$\pm$0.21 for the occurrence rate. This is consistent with the non-imaging study by \cite{Suresh2017} and suitable for coronal heating. 

\citet{Mondal2020b} have also reported slopes steeper than -2 for mSFU level impulsive bursts from the quiet Sun.
We notice an absence of flares below $10^{20}$ ergs in Fig. \ref{Fig:occurance} (B). This could be due to the thresholds set in the section \ref{sec:six_s} for the feature identification and the sensitivity limits of the visibility subtraction approach for the current MWA data.


\cite{Reid2018} studied the properties of the type-III electron beams using simulations, and derived the following expression for the energy of the electron beams ($E_{beam}$) based on the observable parameters:
\begin{equation}
\begin{array}{l}
E_{beam} (erg) = 8.8\times10^{-7} (\frac{\bar{V}}{cm s^{-1}}) (\frac{t_{dur}}{s})  \\ \times (\frac{\theta}{arcmin}) (\frac{f}{Hz})^2 (\frac{S}{SFU})^{0.5}
\end{array}
\end{equation}
where $\bar{V}$, $t_{dur}$, $\theta$, $f$ and $S$ are beam velocity, burst's duration, burst's source size, frequency and source flux density respectively. Using nominal or limited values from our nonthermal bursts, $t_{dur}$ =0.5s, $\theta=3.7'$, $\bar{V}$ = 0.1c, $f=$108 MHz, and $S=$10 milli-SFU,  we obtain energy estimates of $\sim 2\times10^{18}$ ergs, roughly an order of magnitude lower than our estimates (Tab. \ref{tab:2}). 
Given the significant uncertainties in the values of various parameters used, the disagreement is comfortingly low.

\begin{table}[]
    \centering
    \begin{tabular}{|c|c|c|c|}
    \hline
$\nu$ (MHz) & Full Disk & Reg. (1-4) & Fraction Reg.(1-4)\% \\ \hline \hline
108.0  &  176  &  13  &  7 \\ \hline
120.0  &  402  &  19  &  4 \\ \hline
133.0  &  820  &  44  &  5 \\ \hline
145.0  &  1347  &  55  &  4 \\ \hline
179.0  &  2206  &  124  &  5 \\ \hline
197.0  &  2299  &  131  &  5 \\ \hline
217.0  &  3314  &  220  &  6 \\ \hline
240.0  &  2111  &  199  &  9 \\ \hline \hline

Total & 12680 & 805 & 6.3 \\ \hline
    \end{tabular}
    \caption{Number of bursts detected in the full solar disk and specifically from Region 1 to 4 for all frequency bands. The last column is the percentage of bursts originating from Region 1 to 4.  }
    \label{Tab:fraction}
\end{table}

\subsection{Radiated Energy at Radio Wavelengths}
Consider a radio source emitting over a bandwidth $\Delta \nu$ for a time duration $\Delta t$.
An upper limit on the radiation energy emitted \citep[$W$,][]{Elgaroy1977,Ramesh2013,Ramesh2021} during the radio burst can be given as:
\begin{equation}
  W= S_{r,\odot}\ \Omega D^2\ \Delta \nu \Delta t\ e^{\tau_0} ,  
\end{equation} where $S_{r,\odot}$ is the observed flux density of the burst, $D$ is 1 A.U., $\Omega$ is the source solid angle, $\Delta \nu$ and $\Delta t$ are the bandwidth and time duration of the burst respectively. 
The $\tau_0$ is the optical depth at the burst site. For the bursts, we assume $\Delta \nu =$ 5 MHz \citep{Suresh2017} and $\Delta t =$ 0.5 s, as most of the bursts are unresolved in time. The mean flux $S_{r,\odot}$ was obtained from the bursts and PSF-sized compact source. The burst radiation arising from coherent emission mechanisms is likely to have some directivity.

We use angular extent of the emission provided by
\citet{Steinberg1974} of $\Omega \sim 0.15$ steradian.
Table \ref{tab:2} list the values of the $S_{r,\odot}$ and $W$ for all the MWA frequency bands. We note that the flux density levels of the bursts lie in the mSFU range and $W$ in the range of $10^{19}-10^{20}$ ergs, much smaller than the magnetic field and thermal energy. The fraction $\gamma=W/E_{nth}$ is also listed in Table \ref{tab:2} and typically lies between $10^{-6} - 10^{-5}$. We note that the radiated energies are an order of magnitude less than the nonthermal particle energies. 

\subsection{Estimates of Magnetic and Thermal Energy Reservoir} \label{sec:model}
The nonthermal population derives its energy from the available reservoirs of the magnetic field energy.
The magnetic field energy and thermal energy density are given by $B^2/8\pi$ and $n_e k_B T_e$ respectively. In a given volume, the total magnetic field energy (E$_B$) and thermal energy (E$_{th}$) are given as
\begin{equation}
E_B=\frac{B^2}{8\pi}Al  
\end{equation} and 
\begin{equation}
E_{th}=n_e k_B T_e A l,
\end{equation}where $A$ is the area and $l$ the column length of the emitting region. 
Using FORWARD $n_e$, $T_e$ and $|B|$ profiles, we compute E$_B$ and E$_{th}$, which are listed in Table \ref{tab:2}. The thermal energy is an order of magnitude lower than the magnetic field energy.
We note that the coronal magnetic field energy lies at the lower end of observed microflares, i.e between $10^{24}-10^{25}$ ergs (Table \ref{tab:2}). A minuscule fraction of the reservoir of $\sim 10^{-4}$ and $\sim10^{-9}$ is released into nonthermal particle energy and radiation energy respectively. Using numerical modelling of the plasma emission and turbulence, \cite{Volokitin2018} estimates radiation efficiency as $2\times10^{-7}-10^{-8}$ , which is the energy carried by radiation w.r.t energy of the Langmuir waves in the flaring volume.

\section{Discussion} \label{sec:discussion}

This work presents the first application of the visibility subtraction method using a running median filter to sensitive quiet Sun MWA observations.
We successfully demonstrate the ability to identify and locate radio bursts down to mSFU flux density and kilo-Kelvin $T_B$ levels using this technique. The main results are: 
\begin{enumerate}
    \item The weak bursts show nanoflare-like spatio-temporal characteristics -- distributed all over the Sun and impulsive in nature.
    \item In terms of their classification, the detected bursts overlap in their characteristics with type-I and type-III solar radio bursts. 
    \item We also detect what is likely to be the faintest type-III radio burst reported yet and another feature showing quasi-periodic bursts.
\end{enumerate}
Each of these are discussed in more detail in the following text.

\begin{figure*}
    \centering
    \begin{tabular}{ccc}
            \resizebox{50mm}{!}{
    \includegraphics{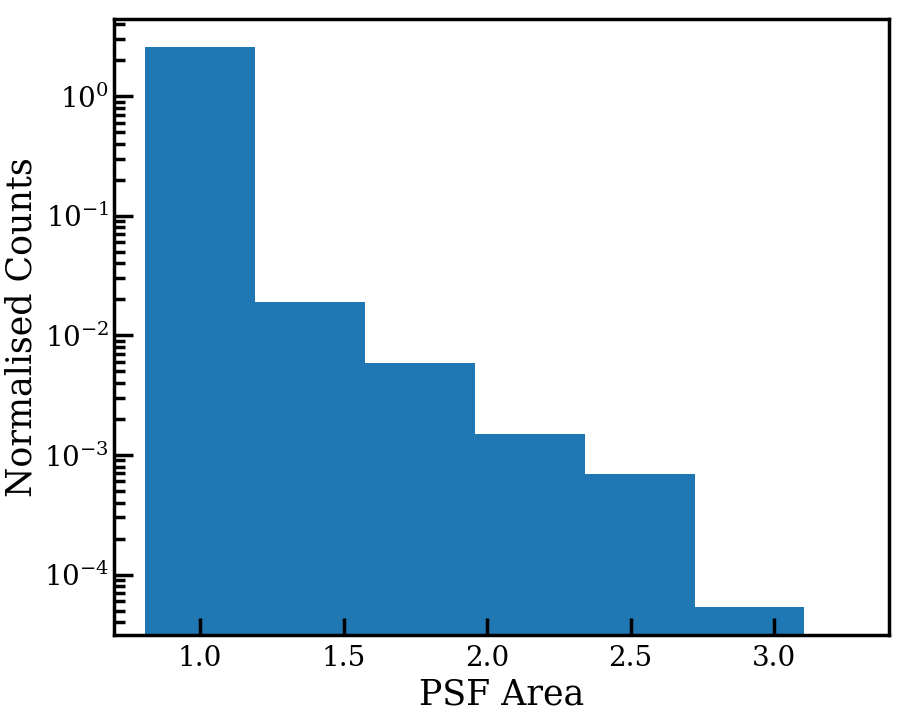}} &
        \resizebox{56mm}{!}{
    \includegraphics{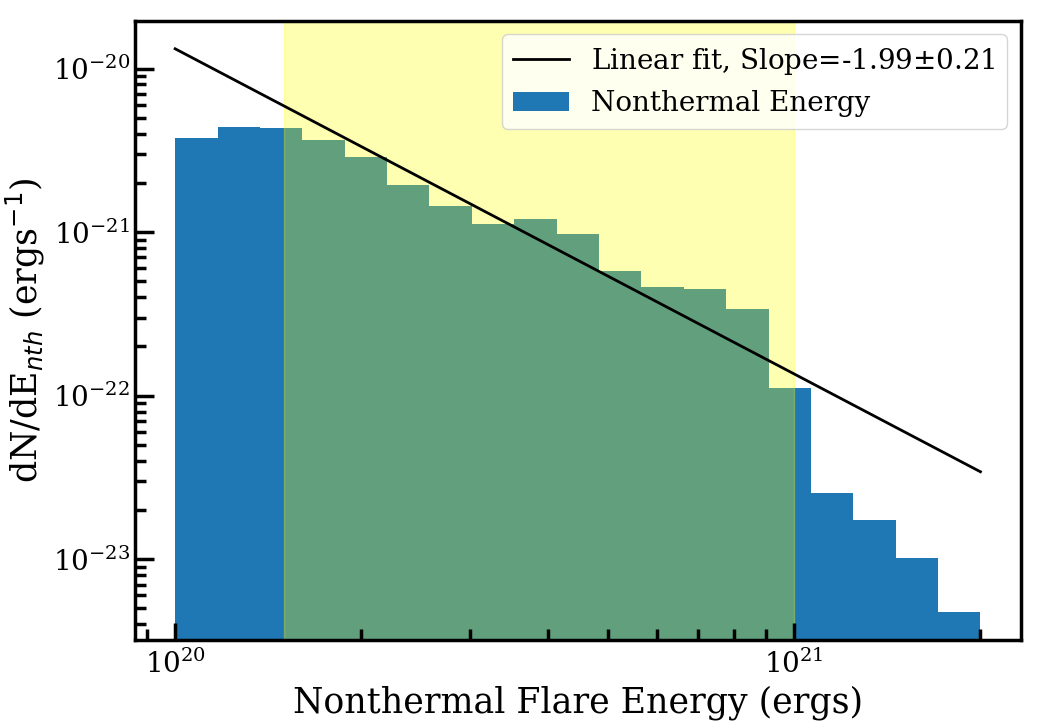}} &
    \resizebox{54mm}{!}{
     \includegraphics{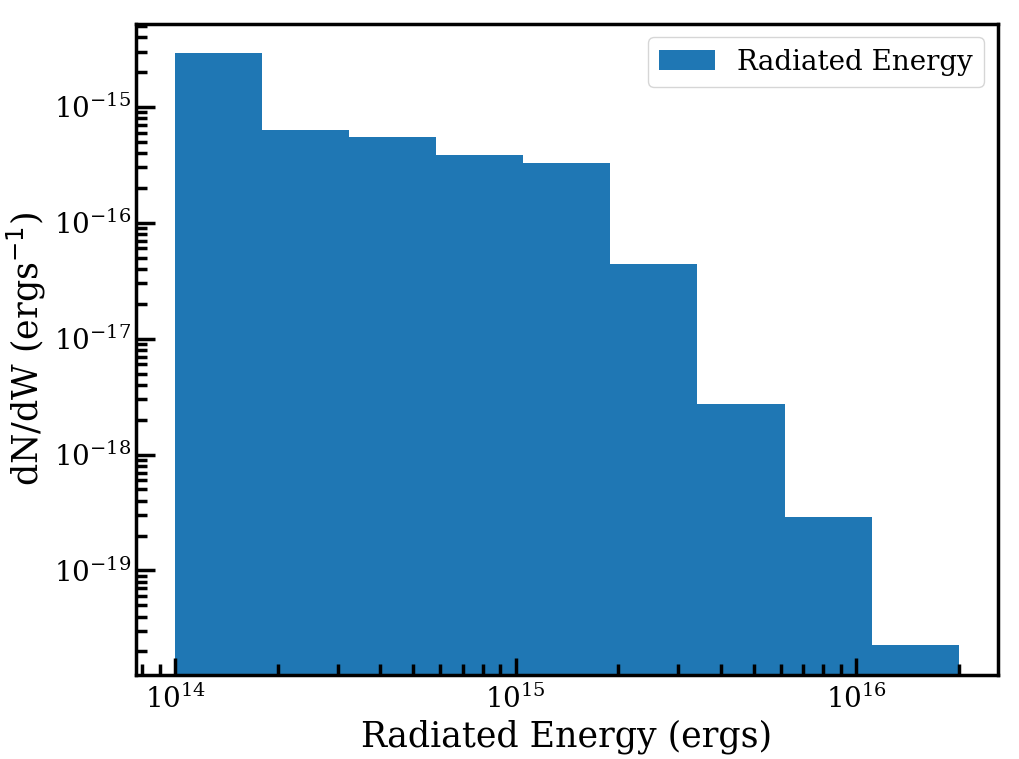}}
    \\ 
(A) Area & (B) Flare Energy Distribution & (C) Radiated Energy distribution
\end{tabular}

    \caption{Properties of the detected nonthermal bursts. Panel A shows the distribution of the area of the bursts relative to the beam area. Note that only 2.5\% events have area greater than a PSF Area. Panel B shows the distribution of the nonthermal particle energies for all the detected radio bursts. Note that the range between $1.5\times10^{20}-10^{21}$ ergs marked in yellow region was used for linear fit in the log-log space. Panel C shows the distribution of the radiated energy of the detected bursts.} 
    \label{Fig:occurance}
\end{figure*}

\subsection{Relationship with nanoflares and coronal heating}

\cite{Parker1988} characterise nanoflares to be localised and impulsive energy releases with $\leq 10^{24}$ ergs. Their energy and spatial distribution are driven by the footpoint motions of the magnetic fields rooted in the photospheric convection occurring all over the Sun. 
The resultant magnetic reconnections are ubiquitously present in the solar corona. The coronal material in the magnetic loops is also dynamic where plasma can move to higher corona from lower heights, sometimes seen as EUV loop top emission in the post-flare times \citep[e.g.][]{Masuda1994,Sharma2016}. 
In higher corona, meterwave non-imaging observations of moderate to quiet Sun using MWA data shows the presence of sub-SFU level nonthermal bursts in the dynamic spectrum \citep{Suresh2017, Sharma2018}. Their flux density distribution during such solar conditions extends to a few to tens of SFU.  \cite{Mondal2020b} detected $T_B$ fluctuations in quiet Sun maps with distributions consistent with a nanoflare based coronal heating scenario down to mSFU levels. 
The estimates of the released energy associated with type I bursts lie in the pico-flare regime using non-imaging observations \citep{Ramesh2013}. 
Here we report mSFU bursts and the distribution of their locations. We determined the locations of $\sim 10^4$ radio bursts with strengths at mSFU levels in half an hour of data at multiple frequencies spanning the range from 108 to 240 MHz (Fig. \ref{Fig:occurance}(A)). 
The robust detection of these bursts in residual maps provides independent confirmation of the presence of these weak and hard to detect features.
We note that the spectral integration used here is an order of magnitude larger than that used by \citet{Mondal2020b} (2 MHz vs 160 kHz).

We interpret the narrowband and impulsive radio bursts as emissions from nonthermal particles interacting with the ambient plasma. The majority of the bursts are not resolved by the 0.5 s instrumental time resolution. 
The associated nonthermal electron energies range from $10^{20}$ to $10^{21}$ ergs, extending into the sub-picoflare category. The flare energy distribution follows a power law with a slope of -1.99$\pm$0.21 (Fig. \ref{Fig:occurance}(B)).
Though this relies on assumptions, it meets the criterion of being relevant for coronal heating \citep{Hudson1991}. 
The regimes over which these power-law slopes have been obtained span about an order of magnitude.
While this span is less than ideal, the power-law slope is a crucial parameter of interest from a coronal heating perspective.
The poor statistics limits us at the higher energy end due to the limited observing duration and the lower energy end by a combination of the imaging quality and instrumental sensitivity. 
It is interesting that power-law lies in the right ball-park for coronal heating, though under the assumptions of uniform optically thick nonthermal plasma emission with typical flare area and column depths.

The requirement of the coronal heating varies from $\approx 10^{6}$ erg cm$^{-2}$ s$^{-1}$ for active regions to $\approx 10^{4}$ erg cm$^{-2}$ s$^{-1}$ for coronal holes \citep{Aschwanden2001}. Using a typical quiet Sun value of $10^{5}$ erg cm$^{-2}$ s$^{-1}$, a back calculation on $\beta$ yields values in the range $10^{-4} - 10^{-5}$. Hence, the assumption of $\beta \approx 10^{-5}$ used in Sec. \ref{sec:energetics} is reasonable.
With the improvement provided by the recently developed MWA solar radio imaging pipeline \citep{Kansabanik2022} and the availability of more sensitive instruments like the MWA Phase III and the future SKA1-Low, this will be a very fruitful area to explore.

We note a more significant occurrence of radio bursts at higher frequencies, i.e. closer to the solar surface. This could be due to the complex topology and higher strength of magnetic fields at lower coronal heights. 
Although the location of the bursts seems uniform, the brighter ones seem to concentrate in a few areas visible in Fig. \ref{Fig:Tb_map}(A). Many of these lie in regions 1 through 4, associated with the bright EUV areas. However, bursts originating near the active regions contribute only about the 4-10\% of the total bursts detected (Tab. \ref{Tab:fraction}). 
The fraction of the bursts in the vicinity of bright EUV features are slightly higher at lower coronal heights by 5--6\%, which is consistent with the standard topology of the coronal loops, i.e. low magnetic strength. 

The locations of the majority of the bursts suggest a mostly uniform distribution with some patches, especially at higher frequencies. However, a few patches show the presence of a more significant number of features (Fig. \ref{Fig:number} (A)). 
We find that the radio bursts are displaced by 3'-8' concerning the associated EUV bright regions. Such displacements are expected at meter wavelengths due to scattering and refraction effects and are routinely seen in MWA data  \citep[e.g.][]{Mulay2019,Sharma2020}. 

With its optical depth in the vicinity of unity, the solar plasma at meter wavelengths lies between the formally optically thick and thin media. 
Hence, the solar limb lines-of-sight sample larger coronal volumes making it more likely to detect a more significant number of bursts near the limb.
Though the presence of scattering effects moderates this, nonetheless, from 145 MHz to 240 MHz, we observed brighter bursts from limb regions in Fig. \ref{Fig:Tb_map} (A). For 217 MHz, the brighter patches capture the North-South curve produced by the solar disk on the eastern and western limbs and have also been noticed in \cite{Sharma2020}.

\subsubsection{Flare Energetics}

The energetically weak flares have been searched for in high-temperature EUV lines, X-ray and radio wavelengths. The weakest X-ray flares have been detected down to sub-A class emission levels by the NuSTAR Observatory \citep{Kuhar2018,Marsh2018,Glesener2020}. The Solar X-ray Monitor (XSM) spectrometer on-board India's Chandrayaan-2 mission regularly detects thermal emission from sub-A class flares \citep{Santosh2021, Ramesh2021}.  
New generation X-ray observatories like the Spectrometer Telescope for Imaging X-rays (STIX) onboard Solar Orbiter also routinely observes sub-A-class flares \citep{Krucker2020,Battaglia2021}. At EUV wavelengths, statistical studies of microflare in quiet regions and flares inactive regions have been performed by, e.g. \citet{Benz2002} and \citet{Lorinck2020}. However, detecting nonthermal particle signatures in these faint flares remains a challenge. 

In radio wavelengths, the detection of the nonthermal population is intrinsically favourable compared to the EUV and X-rays. In microflares, using the 20 SFU gyrosynchroton source, the nonthermal population can be constraint down to 4 keV \citep{Sharma2020M}. The faint radio burst like \cite{Ramesh2013} associated with type-I bursts detected at $\sim 14$ SFU level. The radio bursts presented here are three orders of magnitude fainter than this at approx mSFU levels. For the radio bursts using the total nonthermal electron number and their energies ($10^{20}-10^{21}$ ergs), the nonthermal energy of the electrons is $\approx 6.67\times (10^{-9} - 10^{-10})$ ergs $= 0.4-4.0$ keV.  In comparison to nonthermal electron energies computed from X-ray, EUV and other radio observations, the presented radio bursts represent energetically much lower accelerated population. The ubiquitous occurrence of keV and sub-keV electron populations and their dissipation in the corona can provide a steady heat flux source for the solar corona. Their occurrence at meterwaves suggests the particle acceleration in the upper higher corona at heights between $\sim$15-150 Mm. Particle detection often below one keV from the in-situ instruments like ISEE3 and  WIND 3DP suggests a strong association (99\%) with the radio type-III bursts \citep{Potter1980, Wang2012ApJ}. Therefore, the presented weak bursts possibly relevant for the escaping particles into the interplanetary medium and heliophysics.

\subsubsection{Indications of additional fainter impulsive emissions}

As compared to radio bursts characterised in section \ref{sec:six_s}, we observe hints of a weaker level of impulsive variability in these datasets. This emission is not associated with EUV bright features and is located near and around the central section of the solar disk. Figure \ref{Fig:averaged_map} shows the time-averaged residual maps after removing $6-\sigma$ and above features. This emission is distributed across the solar disk and above or comparable to the self-noise. 
Such an increase is likely due to the presence of energetically small bursts weaker than the threshold criteria defined in Sec. \ref{sec:six_s}.
Features at this level however, lie close to the instrumental noise and one needs data with higher sensitivity to probe these emissions reliably.



\subsection{Comparison with Type-I and III radio bursts}
The weak bursts presented here can be interpreted as either of two well known types of radio bursts: type I or type III. 
Type I bursts originate from trapped particles in magnetic fields, typically in the vicinity of the active regions. The occurrence rate of the flux of the type-I bursts have been shown to vary between -4 to -2 \citep{Mercier1997,Ramesh2013, Iwai2014}. In addition to intrinsic variations, differences in instrumental sensitivity and time and frequency resolution can also contribute to the large variations in the observed occurrence rates. 

Weak type III bursts originating from travelling electron beams are the other candidate for explaining the weak bursts. \cite{Saint-Hilaire2013} found the $T_B$ occurrence rate slope $\approx$-1.8 in meter-wavelengths for type-III in observations spanning a decade. 
We detect a variation in the $T_B$ occurrence rates within meterwavelengths from -0.42 to -2.6. The harder rates dominate in the middle frequency bands 145 MHz to 217 MHz, where there are fewer events greater than $10^5$ K during half an hour of the observation. Even though, we do not resolve most bursts temporally, we detect a drifting feature in frequency, discussed in the next section. 
Overall, the spectro-temporal characteristics of the bursts are consistent with either of weak type I or type III like bursts.


\subsection{Interesting events}

Among detected radio bursts, there are two intriguing faint bursts -- a group of quasi-periodic bursts and a spectrally drifting Type III burst, which are discussed in the following text. 

\subsubsection{Quasi-periodic bursts}


 
The phenomenon of quasi-periodic bursts can be attributed to many physical processes like periodic reconnections, loop oscillations etc. \citep{Nakariakov2009,Kashapova2021}. In meterwaves, several quasi-periodic oscillation during type-I \citep{2021ApJ...920...11M, Mohan2021a} and type-III bursts \citep{Mohan2019} have been reported. We observe a group of narrowband quasi-periodic bursts at 240 MHz (Fig. \ref{Fig:ds_regions} (C) top panel). The bursts have a period of $\sim 20$ seconds. 
To our knowledge, these are weakest bursts of this kind reported so far, with peak $T_B$ reaching only 20 kK. In the total power dynamic spectrum of 240 MHz, we do observe a minute spike of $\sim$ 0.2 SFU shown in Fig. \ref{Fig:intro2} (C) at 03:56 UT. The fractional increase for this spike over the 19 SFU background is $\sim 1\%$. This agrees well with the fractional change in $T_B$, i.e. 20 kK over 2 MK. These bursts occurred in region 1 on the close to the active regions on the western limb. Each burst in the series of quasi-periodic bursts lasts $\sim 6$ sec. The sporadic and impulsive time profiles suggests that they are unlikely due to long-lasting loop oscillation, but more plausibly due to periodic wave-particle interaction or periodic reconnection.
 
\subsubsection{Faintest Type-III burst}
Type III burst are the most commonly occurring radio bursts. In the observed 30 mins, we detect a weak type III burst shown in Fig.~\ref{Fig:ds_regions} (C), bottom panel. The drifting feature starts at 145 MHz and ends at 108 MHz. The $T_B$ increases with decreasing frequency, with peak $T_B$ at $\sim 60$ kK. To our knowledge, this is the faintest type-III burst on record. 

The drift rate of the feature is $\sim 18.5$ MHz/s, which is slower than usual type-III drift rates. Using the coronal densities (Sec. \ref{sec:energetics}), we compute plasma frequency and corresponding coronal height range of the travelling electron beam from 18 Mm to 63 Mm with a beam speed of $\sim$0.1$c$. Such speeds are very similar to the $\sim0.3$c routinely seen in brighter type-III bursts \citep{Saint-Hilaire2013,McCauley2017}.
The presence of weak shock driving quiet Sun type-III have been simulated by \cite{Li2012}. The weak shocks can trigger the faint type-III bursts with $< 1$ SFU. Also, the increase in apparent angular size due to the strong scattering can decrease the observed brightness temperature.





\section{Conclusion} \label{sec:conclusion}

This work presents the detection, imaging and characterisation of mSFU level impulsive compact emissions ubiquitous on the Sun even during fairly quiet times at meterwaves using the continuum visibility subtraction technique.
They are found to be distributed essentially uniformly all over the solar disk with a few clusters of higher occurrence rates.
We estimate the burst energies to lie in the range $\sim 10^{19}-10^{21}$ ergs, in the sub-pico flare range, and the electron energies in the range 0.4-4.0 keV.
Though their overall spatial distribution is rather smooth, the more energetic of these bursts come from the vicinity of the active regions and the EUV bright regions.
Their $T_B$ lies in the range of a few to many kK.
We also report the faintest quasi-periodic radio burst (peak $T_B \sim 20$kK) and the faintest type III (peak $T_B \sim$60 kK).

Similar features have been reported recently by \citet{Mondal2020b}, who used much smaller bandwidths, and interpreted them as the radio signatures of weak particle acceleration events arising due to magnetic reconnections believed to take place everywhere on the Sun, as hypothesised by \citet{Parker1988}.
The technique of imaging using residual visibilities, employed here for the first time on MWA data, presents a significant improvement over earlier work by providing images of these weak emissions and an independent confirmation of existence of such bursts.
Given the potential implications of this discovery, in context of the coronal heating problem, and the fact that the weak transient nature of these emissions requires one to push the data close to its limits, it is essential to examine these data in multiple different ways using independent algorithms and pipelines to firmly establish the reality of these weak bursts. This work is an important step in this direction and points to presence of reconnection based heating processes operating in the 15-150 Mm coronal height range. 
Our work also suggests the presence of an even weaker population of bursts, though the investigation of those will need to wait for more sensitive data. The future SKA-Low will be a very promising instrument for such studies.


\section*{Acknowledgments}
{This scientific work makes use of the Murchison Radio-astronomy Observatory (MRO), operated by the Commonwealth Scientific and Industrial Research Organisation (CSIRO).
We acknowledge the Wajarri Yamatji people as the traditional owners of the Observatory site. 
Support for the operation of the MWA is provided by the Australian Government's National Collaborative Research
Infrastructure Strategy (NCRIS), under a contract to Curtin University administered by Astronomy Australia Limited. We acknowledge the Pawsey Supercomputing Centre, which is supported by the Western Australian and Australian Governments. 
This research has also made use of NASA's Astrophysics Data System (ADS). 
RS acknowledges support of the Swiss National Foundation, under grant 200021\_175832. RS acknowledge Dr. Bin Chen, NJIT and Dr. Sijie Yu, NJIT for useful discussions.
DO acknowledges support of the Department of Atomic Energy, Government of India, under the project no. 12-R\&D-TFR-5.02-0700.}



\bibliography{manuscript}
\bibliographystyle{aasjournal}


\end{document}